\documentclass[reprint,NumberedRefs]{JASA}
\usepackage{siunitx}
\usepackage{filecontents}
\usepackage{rotating}
\usepackage{multirow}

\begin{document}
% PREPRINT: 
\title[Expert decision support system]{Expert Decision Support System for aeroacoustic source type identification using clustering}

\author{A. Goudarzi}
\email{armin.goudarzi@dlr.de}
\author{C. Spehr}
\email{carsten.spehr@dlr.de}
\affiliation{German Aerospace Center (DLR), Germany}

\author{S. Herbold}
\email{herbold@cs.uni-goettingen.de}
\affiliation{Institute of Computer Science, University of G\"ottingen, Germany}

\begin{abstract}
This paper presents an Expert Decision Support System for the identification of time-invariant, aeroacoustic source types. The system comprises two steps: first, acoustic properties are calculated based on spectral and spatial information. Second, clustering is performed based on these properties. The clustering aims at helping and guiding an expert for quick identification of different source types, providing an understanding of how sources differ. This supports the expert in determining similar or atypical behavior. A variety of features are proposed for capturing the characteristics of the sources. These features represent aeroacoustic properties that can be interpreted by both the machine and by experts. The features are independent of the absolute Mach number which enables the proposed method to cluster data measured at different flow configurations. The method is evaluated on deconvolved beamforming data from two scaled airframe half-model measurements. For this exemplary data, the proposed support system method results in clusters that \added[R1C40]{mostly} correspond to the source types identified by the authors. The clustering also provides the mean feature values and the cluster hierarchy for each cluster and for each cluster member a clustering confidence. This additional information makes the results transparent and allows the expert to understand the clustering choices.
\end{abstract}

\maketitle

\section{Introduction}\label{sec:introduction}

Multiple noise-generating phenomena and mechanisms exist in aeroacoustics~\citep{Mueller1979, Howe2007}. To identify these phenomena in processed experimental data, expert domain knowledge and a detailed analysis of measurements are necessary. For the localization and estimation of the sound power of complex source geometries, such as planes, cars, or trains, beamforming is a reliable method~\citep{Beamforming2019}. Beamforming results in high-dimensional maps of the Power Spectral Density (PSD) that are difficult and time-consuming to analyze manually. Therefore, the resulting beamforming maps are usually integrated over space and frequency areas to obtain low-dimensional data such as spectra~\citep{Martinez2019} or Overall Sound Pressure Levels (OASPL) which human experts then analyze and interpret. For the identification of the source types or source mechanisms experts often rely upon the spatial localization and meta-information, e.g., that sources are located at a trailing edge and thus, are identified as trailing edge noise. For real-world vehicles or vehicle models, this information can be missing or misleading, since the geometries of interest are highly complex and result in a superposition of sound generating mechanisms. Therefore, identifying aeroacoustic source types is a complex, time-consuming task that requires experts to compare the analyzed data to simplified, generic, and controlled reference measurements. However, these reference measurements often do not reflect the behavior of real-world geometries due to mismatched Reynolds numbers, installation effects, and object modifications (e.g. tripping, and empty nacelles in airframe models).\\

The goal of this paper is to overcome this obstacle by introducing a system, that supports experts in the process of source identification by means of automated data processing and machine learning.\deleted[R2C04a, R2C42]{ Machine learning is a powerful tool that currently outperforms any other automated classification system given enough learning data~\hbox{\citep{Mello2018, Bianco2019}}.} There are two major categories for analyzing and structuring data: supervised learning and unsupervised learning. Supervised learning models an unknown function\deleted[R2C42]{ $y=f(x)$} for known input data\deleted[R2C42]{ $x$} and a desired outcome\deleted[R2C42]{ $y$}. The desired outcome\deleted[R2C42]{ $y$} must be known, i.e., the data must be labeled with a ground truth\added{~\hbox{\citep{Mello2018, Bianco2019}}}. \deleted[R2C04b]{Examples of successful application of supervised learning to classify sounds can be found in many fields, including bioacoustics~\hbox{\citep{Skowronski2006,Shamir2014,Flaspohler2017,Li2019,Hamed2019,Koutsomitopoulos2020}}, aeroacoustics~\hbox{\citep{Cook2020}}, architectural acoustics~\hbox{\citep{Choi2019}}, structural analysis via emitted sound~\hbox{\citep{Regan2017}}, and a number of natural language processing applications, such as mispronunciation detection~\hbox{\citep{Chen2009}}, speech reconstruction from articulatory sensor data~\hbox{\citep{Gonzalez2017}}, the estimation of the vocal fold physiology~\hbox{\citep{Zhang2020}}, or speech~\hbox{\citep{Reddy2004}} and speech emotion recognition~\hbox{\citep{Lotfidereshgi2017,Li2018}}. Based on the use case, the sounds $x$ are processed as raw time-data (if the time structure of the data is important), spectra (for time-invariant sources, or if the acoustic power over frequency is important), or spectrograms and wavelets (if both, time and frequency content are of interest).} \added{However, }the aeroacoustics of airframe noise and complex aircraft models is a field where it is not feasible at the moment to obtain a ground truth due to the complexity of sound generating mechanisms, such as turbulence-induced noise at high Reynolds numbers. This can be accounted to the observations which differ on the sound generating side (e.g., turbulence-induced pressure fluctuations, which are often studied using Computational Fluid Dynamics, Particle Image Velocimetry, Schlieren Imaging, or Rayleigh Scattering), from sound radiation (e.g., due to coherent structures in these fluctuations), and from far-field observations. For seemingly simple mechanisms like the shear layer induced noise of a jet, there exists no scientific consensus on the exact mechanism that is responsible for the observed far-field sound~\citep{Karabasov2010}. Additionally, there is not yet enough data of complex, real-world aircraft models to employ supervised machine learning, since the wind tunnel measurements are costly and both, model geometries and results are often confidential. Therefore, supervised learning is currently not a suitable approach for the presented task.\\

Instead, unsupervised learning will be used in this paper to predict \deleted{clusters of }similar source types or source mechanisms\added{, which is called clustering}. While these clusters cannot replace the manual analysis of the expert, they are supposed to help with identifying similar sound source types, interpreting their acoustic properties, and detecting typical and anomalous behavior. Such a method is referred to as an Expert Decision Support System (EDSS). \deleted{Unsupervised learning does not require labeled data, i.e. a ground truth. Instead, the goal of unsupervised learning is to group the presented data based on its inner structure, which is called clustering. Examples for the use of unsupervised learning are fields where no ground truth can be obtained or where it is practically not feasible to do so.} Bioacoustics in coral reefs~\citep{Ozanich2021} and the structural analysis of wind turbine bearings via their emitted sound~\citep{Benali2018} are examples for unsupervised learning, as the task of labeling requires either knowledge that is unavailable or too much time and resources (e.g., the turbines would have to be dissembled to obtain a ground truth). \deleted[R2C04b]{Similar to supervised learning}The sounds can be processed and presented to the clustering algorithm in many forms. However, since unsupervised learning not only models the function \deleted[R2C42]{$f(x)$} but also estimates an outcome\deleted[R2C42]{ $y$}, it is important that the variance within the data representation makes the desired clustering choice likely. E.g., to successfully cluster sounds of different animals, the data variance between the animals must be greater than the variance \replaced[R1C13]{between}{than the variance due to changes between} the locations the sounds were recorded at. Otherwise, the clustering will result in clusters of different recording locations instead of different animals. This is achieved through data preprocessing, which includes the reduction of the data's dimensionality (e.g., the Overall Sound Pressure Level (OASPL) of a sound instead of time-domain data), and the reduction of \deleted[R1C04c]{information and }complexity (e.g., using the Power Spectral Density\deleted{ (PSD)} instead of the time-data). This preprocessing step and careful selection and definition of calculated properties with the goal to obtain a representation of the underlying data suitable for the clustering is called ``feature engineering''.\\

\replaced[R2C43]{In order to be able to cluster aeroacoustic source types, the source information has to be condensed and transformed to a machine-interpretable data space to meet the requirements described above. To do so, we identify and define features that formalize expert domain knowledge of aeroacoustic properties to enable their automated calculation. }{The paper is structured as follows.} We use CLEAN-SC~\citep{Sijtsma2007} beamforming maps~\citep{Beamforming2019} of the scaled air-frame models of a Dornier 728 (Do728)~\citep{Ahlefeldt2013} and an Airbus A320 (A320)~\citep{Ahlefeldt2017} as example data, featuring multiple aeroacoustic source types. We employ the Source Identification based on spatial Normal Distributions (SIND)~\citep{Goudarzi2021} approach to identify individual sources and obtain their spectra from the beamforming maps. We explain typical aeroacoustic properties and derive corresponding features, discuss their usefulness, and \replaced{specify a proof-of-concept implementation}{propose mathematical definitions}. We then cluster the sources in the obtained feature space using HDBSCAN~\citep{Campello2013}. We present a manual identification of the airframe source types with exemplary spectra and our decision choices to the reader, which allows us to compare the resulting clusters to our source categories. We then evaluate which clusters are meaningful and correspond to our source categories, derive a corresponding confusion matrix, and calculate a clustering accuracy based on them.\\

The method reported in this paper was originally presented at the AIAA Aviation 2020 conference~\citep{Goudarzi2020} as work-in-progress. This paper presents more data (five different Reynolds numbers for the Do728, and four for the A320 versus one for both models in the conference paper). It presents additional features, and some modified feature calculations, an in-depth analysis of the proposed features, and a statistical analysis and discussion of the clustering results.

\section{Datasets}\label{sec:datasets}
The data used in the present paper consists of beamforming measurements of two closed-section wind tunnel models: one is of a Do728~\citep{Ahlefeldt2013} and one is of an A320~\citep{Ahlefeldt2017}. \replaced[R1C02]{Both models were observed at various Reynolds number configurations. For each Reynolds number configuration, several angles of attack $\alpha$ and for each angle of attack several Mach numbers $M$ were examined.}{Both models were observed at multiple Reynolds numbers $\langle \text{Re}\rangle_M$, angles of attack $\alpha$, and Mach numbers $M$.} For the Do728 model, the Mach-averaged Reynolds numbers $\langle\text{Re}\rangle_M$, the ambient pressures $p_0$ and cryogenic temperatures $T$ are shown in Table~\ref{tab:Do728} based on the mean aerodynamic cord length $D_0=\SI{0.353}{\metre}$. \added{With the dynamic viscosity $\mu(T)$ and density $\varrho(p_0,T)$ of the medium, the Reynolds number is
\begin{equation}\label{eq:Reynoldsnumber}
    \text{Re}=\frac{\varrho(p_0,T)M(T,u)D_0}{\mu(T)} \,.
\end{equation}}
Values of $\alpha_a=[\SI{1}{\degree}$, $\SI{3}{\degree}$, $\SI{5}{\degree}$, $\SI{6}{\degree}$, $\SI{7}{\degree}$, $\SI{8}{\degree}$, $\SI{9}{\degree}$, $\SI{10}{\degree}]$ were chosen for angle of attack for every Reynolds number configuration and $M_j=[0.125$, $0.150$, $0.175$, $0.200$, $0.225$, $0.250]$ as Mach number for every angle of attack. In total, the Do728 dataset contains \added[R1C02]{$5\langle\text{Re}\rangle_M\times8\alpha\times6 M$ =} 240 different flow configurations. The array consisted of 144 microphones at an oval aperture of $\SI{1.756}{\metre}\times\SI{1.3}{\metre}$ and a data sample frequency of $f_S=\SI{120}{\kilo\hertz}$ was used.\\
\begin{table}%[ht!]
\caption{\label{tab:Do728}Reynolds configurations of the Do728 dataset.}
\begin{ruledtabular}
\begin{tabular}{r|ccccc}
configuration & D1& D2& D3& D4& D5\\
\hline
$\langle$Re$\rangle_M [\SI{1e6}{}]$  & 1.4& 1.8& 2.5& 3.8&10.6\\
$T [\SI{}{\kelvin}]$ & 290& 250& 200& 150& 100\\
$p_0 [\SI{1e5}{\pascal}]$ & 1.0& 1.0& 1.0& 1.0& 1.0\\
\end{tabular}
\end{ruledtabular}
\end{table}

The A320 model was observed at $\alpha_a=[\SI{3}{\degree}$, $\SI{7}{\degree}$, $\SI{7.15}{\degree}$,$\SI{9}{\degree}]$ for every Reynolds number configuration, and $M_j=[0.175$, $0.200$, $0.225]$ for every angle of attack. The \replaced[R1C02]{mean}{Mach averaged} Reynolds numbers, the ambient pressures $p_0$ and cryogenic temperatures $T$ are shown in Table~\ref{tab:A320} based on $D_0=\SI{0.353}{\metre}$. In total, the A320 dataset contains 48 different flow configurations. The array consisted of 96 microphones at an aperture of $\SI{1.06}{\metre}\times\SI{0.5704}{\metre}$ and the data was recorded at $f_S=\SI{150}{\kilo\hertz}$. \added[R2C44]{For both datasets} the Cross-Spectral density Matrices (CSM) were calculated using Welch's method with a block size of 1024 samples and $\SI{50}{\percent}$ overlap which resulted in around 7000 block averages for the Do728 and 9000 averages for the A320. The beamforming was performed using conventional beamforming~\citep{Beamforming2019} and CLEAN-SC deconvolution~\citep{Sijtsma2007} on a regular grid with a focus point resolution of $\Delta x_1 = \Delta x_2 = \SI{5e-3}{\metre}$. The focus plane for both models is around $\Delta x_3\approx\SI{1}{\metre}$ away from the array and covers around $\SI{2}{\metre\squared}$, which results in a total of \SI{8e5}{} focus points per beamforming map.

\begin{table}%[ht!]
\caption{\label{tab:A320}Reynolds configurations of the A320 dataset.}
\begin{ruledtabular}
\begin{tabular}{r|cccc}
configuration & A1 & A2& A3 & A4\\
\hline
$\langle$Re$\rangle_M [\SI{1e6}{}]$ &1.4 & 5.1& 5.1& 19.9\\
$T [\SI{}{\kelvin}]$ & 310& 311& 125& 120\\
$p_0 [\SI{1e5}{\pascal}]$ & 1.10& 3.99& 1.15& 4.19\\
\end{tabular}
\end{ruledtabular}
\end{table}

\section{Methodology}\label{sec:methodology}
This section presents the methodology of the \replaced[R1C15]{EDSS}{Expert Decision Support System (EDSS)}. First, the procedure is presented and compared to a manual source analysis. Second, the definition of an aeroacoustic source in this context is provided. Third, aeroacoustic properties are discussed and corresponding features are derived. Forth, the clustering process based on the features is described. \replaced[R2C06]{We will use Einstein notation throughout the paper to indicate the dimensionality of the variables.}{We will use italic indices to to indicate the dimensionality of the variables. For averaging a variable $v$ over its $i$-th dimension, we use $\langle v_i\rangle_i$
\begin{equation}
    \langle v_i\rangle_i = \frac{\sum_i^I v_i}{I}\,,
\end{equation}
and for the corresponding standard deviation $\sigma_i(v_i)$
\begin{equation}
    \sigma_i(v_i) = \sqrt{\frac{\sum_i^I (v_i-\langle v_i\rangle_i)^2}{I}}\,.
\end{equation}
}

\subsection{Procedure}
\begin{figure}
	\centering
	\includegraphics[width=\reprintcolumnwidth]{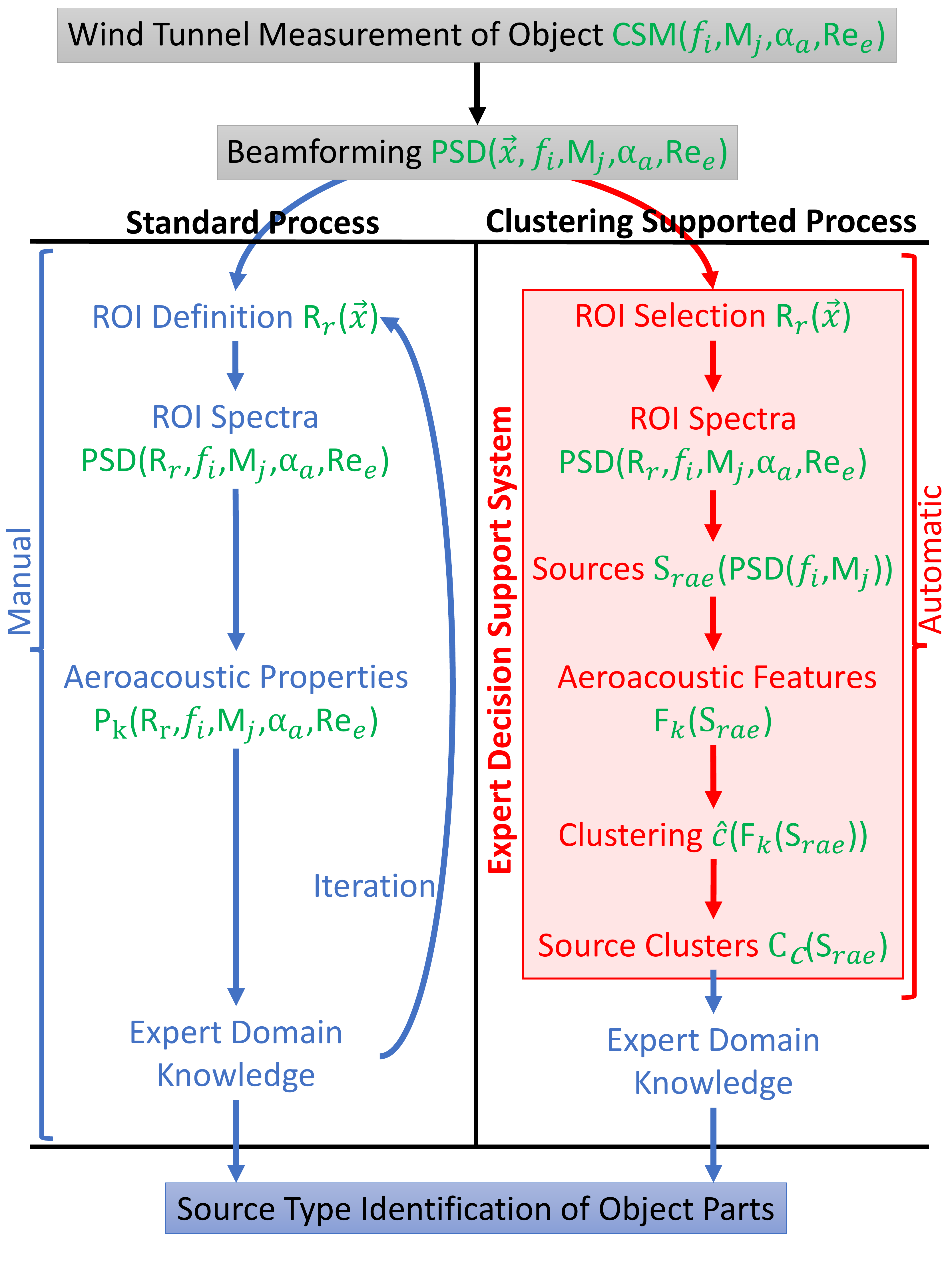}
	\caption{(Color online) Comparison of the evaluation process of wind tunnel beamforming measurements. Left: a standard process using manual analysis. Right: the proposed Expert Decision Support System. Manual processing steps are displayed in blue while automatic steps are displayed in red. The resulting variables of each step are displayed in green. The indices indicate the dimensionality of these variables.}
	\label{fig:Figure1}
\end{figure}

Figure~\ref{fig:Figure1} compares the proposed EDSS process to a standard manual source analysis process for beamforming maps PSD$(\vec{x},f_i,M_j,\alpha_a,\text{Re}_e)$, which typically includes spatial variables $\vec{x}$, multiple frequencies or frequency intervals $f_i$, angles of attack $\alpha_a$, Mach numbers $M_j$, and Reynolds numbers Re$_e$. In the manual process, multiple Regions Of Interest (ROI) $R_r(\vec{x})$ are defined and spatially integrated to derive acoustic spectra PSD$(R_r,f_i,M_j,\alpha_a,\text{Re}_e)$. This first step is already a challenging task since the ROIs must only contain the individual sources to obtain individual source spectra, which can be only verified using the resulting spectra. This often requires several iterations, as described in the following. After definition, the ROI spectra are analyzed which often requires expert knowledge and intuition. Based on the analysis the ROI are then redefined (e.g., if two separate sources are detected within a ROI). Based on the spectra at different Mach numbers, expert knowledge, intuition, and meta-information (e.g., the source is located at a trailing edge and thus, must be trailing edge noise) aeroacoustic properties are derived, and the source type is then identified or vice versa. The main challenge for this process is the high dimensionality of the properties $P_k(R_r,f_i,M_j,\alpha_a,\text{Re}_e)$ and the requirement for an iterative approach.\\

In comparison, the EDSS aims at automating most of these tasks. First, the ROI definition $R(\vec{x})$ and spectra generation were shown to have the capacity to be automated using the SIND method~\citep{Goudarzi2021}. The EDSS then defines a \added{source} $S_{rae}$ for each ROI $R_r$\added[R1C16]{,} at each angle of attack $\alpha_a$\added[R1C17]{, and} at each Reynolds number Re$_e$. This results in \replaced[R2C07]{a PSD$(f_i,M_j)$}{multiple sources $S_{rae}(\text{PSD}(f_i,M_j))$}\deleted{ for each source}, for which aeroacoustic features $F_k(S_{rae})$ are then calculated. Finally, the sources are clustered $\hat{c}(F_k)$ based on this $k$-dimensional feature space which results in a cluster prediction for each source $C_c(S_{rae})$. This provides $c$-dimensional prediction information (one cluster\deleted[R1C18]{ cluster} is predicted for each source). Also, additional information about the similarity of source groups based on their cluster-averaged features is provided. The source types can then be identified manually based on the expert's knowledge and the low-dimensional information provided by the EDSS.

\subsection{Source definition}
An aeroacoustic source emerges either from the interaction of an obstacle placed in a flow, e.g., a cylinder~\citep{Zdravkovich1997} or from the flow itself, e.g., jet noise~\citep{Lighthill1952}. When we observe a source at a specific flow speed or angle of attack, the observed source is often a realization of one or more underlying physical mechanisms. \replaced[R2C15]{These mechanisms often depend monotonously on parameters such as Mach or Reynolds number.}{These mechanisms often result in observations that change continuously over variations of flow parameters such as the Mach or Reynolds number.} To distinguish between different source mechanisms, we have to identify the underlying parameters for which the source mechanism is self-similar. The observable variables depend on the measurement matrix in controlled wind tunnel experiments. They can be the angle of attack $\alpha$, wind speed $u$, Mach number $M$, temperature $T$, and ambient pressure $p_0$. The variation in pressure and temperature changes the Reynolds number, the variation of \replaced[R1C19]{$u$}{the wind speed} and \replaced[R1C19]{$T$}{the temperature} changes the Mach number\deleted[R1C19]{$M$}. Wind tunnel facilities often use scaled models, increased total pressures, and a lowered temperature to achieve high, real-world Reynolds numbers~\citep{Ahlefeldt2013}. \replaced[R1C20]{For these flow variations, the source's ability to radiate acoustic energy to the far-field can be observed using beamforming, which is described by the source emission power quantified by the Power Spectral Density (PSD)~\hbox{\citep{Beamforming2019}}.}{Thus,} the result of beamforming in wind tunnels is a high-dimensional sound power map $\text{PSD}\left(\vec{x},f,M(T,u),\alpha,\text{Re}(T,u,p_0)\right)$.\\

A basic assumption in aeroacoustic source modeling is that small changes in subsonic Mach numbers do not alter the source mechanism~\citep{Howe2007}. Also, a source may exist in extended spatial areas or frequency intervals, for instance, the sound generated from vertices in Kármán's vortex street~\citep{Prandtl1957,Zdravkovich1997}. Sources may have a frequency-dependent spatial location and shift their peak frequencies with \replaced[R1C21]{increasing}{changing} Mach number. One example is jet noise where the location of dominant sound generation shifts downstream and the peak frequency decreases with increasing Mach number, while the PSD level increases~\citep{Lighthill1952}.\deleted[R1C22]{Increasing the Mach number can therefore cause a change in sound power or a shift in peak frequencies and source location.} However, a normalization of the PSD or frequency using the Mach number reveals that the spectrum shape, Mach-normalized peak frequencies, and Mach-normalized PSDs collapse~\added[R2C46]{\hbox{\citep{Quinlan1996}}}. This is referred to as self-similarity. The underlying physical sound-generating mechanism has not changed, we only observe a different realization of the source mechanism. However, source mechanisms can change completely for large variations of their Reynolds numbers. One example for this is the radiated acoustics from a flow around a cylinder~\citep{Zdravkovich1997} where the flow properties may change due to, for instance, the transition from a laminar to a turbulent flow. \added[R1C26]{Thus, only small subsonic Mach number changes are considered. }All other variations such as the angle of attack $\alpha$ or changes in geometry may alter the dominant source mechanism abruptly (e.g., slat tones and flow separation in airframe noise may appear). We treat these variations as potentially different source mechanisms. \\

Since we are interested in clustering the sources according to their underlying physical mechanisms rather than clustering them by their dominance, we need to analyze the scaling behavior over Mach number and the similarity laws~\citep{Howe2007} rather than absolute levels at specific Mach numbers. Therefore, we define \replaced[R1C23, R2C47]{that a source with the following properties}{the properties of a source as follows}. First, a source is connected to a spatial region $R(\vec{x})$. \replaced[R1C24,R1C25]{Second, a source is defined for a single Reynolds number configuration and angle of attack. A large change in Reynolds number results in a different source. Third, a source can be observed at different Mach number, it is still the same source.}{Second, when a source is observed at different Mach number, it is still the same source. Third, a source may be observed for small variations of the Reynolds number due to the change of the Mach number. A large variation of the Reynolds number to to changes in temperature or pressure results in a different source. Forth, any change of the angle of attack results in a different source.} Based on this definition a sound source's PSD, obtained from the spatially integrated ROI, possesses the free variables $\text{PSD}\left(f,M\right)$ at a fixed $\langle\text{Re}\rangle_{M}$ and $\alpha$. For example, a global ROI $R(\vec{x})$ that was identified by SIND within the Do728 beamforming maps is treated as unique sources for each angle of attack and \added{Mach averaged} Reynolds number (which results in $8\alpha\times5\langle\text{Re}\rangle_M=40$ individual sources for any identified spatial ROI), each represented by six spectra at different Mach numbers $M=[0.125,\dots,0.250]$. Since we assign multiple spectra of measurements at different Mach numbers to one source, we can derive its acoustic properties not only from the individual spectra but from the changes over Mach number or as an average property of the spectra. This has the advantage, that the Mach-normalized features of different datasets are comparable despite their measurement at different Mach numbers. Additionally, the averaging reduces relative errors.\\

\deleted[R1C26]{This source definition has two limitations. First, the aeroacoustic properties depend on the Reynolds number, which itself linearly depends on the flow speed. As discussed above, a change in Reynolds number may change the source mechanism. Often, this is not a problem since the acoustic properties do not change significantly over multiple orders of magnitude of the Reynolds number~\hbox{\citep{Howe2007}}. As we show in Section~\ref{sec:mod_exponent}, the variation of the Reynolds number can result in a drift of the Strouhal number with increasing Mach number. We introduce a normalized frequency which is a generalization of the Strouhal and Helmholtz number, which accounts for this drift. Consequently, we neglect the change in Reynolds number caused by the change in Mach number and assume a Mach-averaged mean Reynolds number for the sources. Second, the source mechanisms can also vary fundamentally with Mach number, especially under trans- and supersonic conditions, and even at constant Reynolds number (e.g. jet screech~\hbox{\citep{Raman1999}}). Therefore, we focus on small, subsonic Mach number changes. Still, the source spectra presented in this paper suggest that many source mechanisms exhibit a small dependency on subsonic Mach number changes, which cannot be captured with the proposed method.}

\subsection{Feature engineering}\label{sec:features}
We require a set of features that describe the aeroacoustic properties of a source for clustering. \deleted[R2C04e]{The process of identifying these and formalizing these properties is referred to as feature engineering. }For optimal clustering results and interpretability of the results we require a feature-set that meets the following conditions:
\begin{itemize}
\itemsep-0.1em 
	\item All features together must unambiguously describe a source or its mechanism.
	\item A feature must describe a basic property of a source and must provide additional information.
	\item The calculation of a feature must be automatable and robust
	\item A feature must be represented by a single numerical value.
	\item A feature must correspond to a physical property.
\end{itemize}
In real-world applications, it is typically not possible to fulfill these requirements completely. Additionally, it can only be analyzed how well the introduced features meet these requirements in the context of the observed sources. In the following we identify aeroacoustic properties and brake them down to numerical features in the subsequent sections. \added[R2C37]{A complete list of the features is given in table~\ref{tab:features_table}.} We will present resulting feature values in the results section and discuss how well the proposed features meet the conditions in the discussion section.
%We identified the following aeroacoustic properties which are broken down to numerical features in the subsequent sections:
% As the procedure is intended to support the expert, the last condition is necessary to enable the interpretation of the results. 
% \begin{itemize}
% \itemsep-0.1em 
% 	\item The broadband self-similarity, see sub-section~\ref{sec:similarity}.
%     \item The frequency normalization exponent, see sub-section~\ref{sec:mod_exponent}.
% 	\item The sound power scaling, see sub-section~\ref{sec:Mach_number_scaling}.
% 	\item The tonality, see sub-section~\ref{sec:tonality}.
% 	\item The source location dependency on the Mach number, see sub-section~\ref{sec:srs_mov}.
% 	\item The spatial source distribution, see sub-section~\ref{sec:srs_distribution}.
% 	\item The spectrum shape, see sub-section~\ref{sec:spectrum_shape}.
% \end{itemize}

\added[R2C37]{
\begin{table}%[ht!]
\caption{\label{tab:features_table}Table of all aeroacoustic properties, their corresponding features, their variables,  equations, if they are used logarithmically with $\log(|v|+1)$, and their (log) value range.}
\begin{ruledtabular}
\begin{tabular}{cc|cccc}
property & feature & var. & eq. & log & range\\ 
\hline
\multirow{3}{*}{\parbox{1.2cm}{self-similarity}} & scal. over St number& scal(St)& \ref{eq:scalability}&no&$[0,1]$\\
 & scal. over He number& scal(He)& \ref{eq:scalability}&no&$[0,1]$\\
 & freq. norm. exp. & $m^\star$ & \ref{eq:m_star} & no & $[0,\infty[$\\
\hline
\multirow{2}{*}{\parbox{1.2cm}{power scaling}} & M scal(St)&$n$&\ref{eq:Mach_scaling} &no&$[0,\infty[$\\
& M scal(He)&$n$& \ref{eq:Mach_scaling}&no&$[0,\infty[$\\
\hline
\multirow{14}{*}{tonality} &number of tones&$\hat{P}_\text{n}$&\ref{eq:tonality_number}&yes&$[0,\infty[$\\
 &tone St shape&$k_\text{St}$&\ref{eq:tonality_gamma}&yes&$[0,\infty[$\\
 &tone St scale&$\theta_\text{St}$&\ref{eq:tonality_gamma}&yes&$[0,\infty[$\\
 &tone St loc  &$l_\text{St}$&\ref{eq:tonality_gamma}&yes&$[0,\infty[$\\
 &tone width shape&$k_\text{w}$&\ref{eq:tonality_gamma}&yes&$[0,\infty[$\\
 &tone width scale&$\theta_\text{w}$&\ref{eq:tonality_gamma}&yes&$[0,\infty[$\\
 &tone width loc&$l_\text{w}$&\ref{eq:tonality_gamma}&yes&$[0,\infty[$\\
 &tone prom shape&$k_\text{p}$&\ref{eq:tonality_gamma}&yes&$[0,\infty[$\\
 &tone prom scale&$\theta_\text{p}$&\ref{eq:tonality_gamma}&yes&$[0,\infty[$\\
 &tone prom loc&$l_\text{p}$&\ref{eq:tonality_gamma}&no&$[0,\infty[$\\
 &scal. over St number&$\text{scal}_\text{p}$(St)&\ref{eq:tonality_scal_p}&no&$[0,1]$\\
 &scal. over He number&$\text{scal}_\text{p}$(He)&\ref{eq:tonality_scal_p}&no&$[0,1]$\\
 &tone intensity&$\text{prop}_\text{p}$&\ref{eq:tonality_prop_p}&no&$[0,1]$\\
\hline
source loc. &source movement&$\Delta
l$&\ref{eq:srs_mov}&no&$[0,\infty[$\\
\hline
\multirow{2}{*}{\parbox{1.2cm}{spatial dist.}} & source compactness&$A$&\ref{eq:source_dist_A}&no&$[0,\infty[$\\
 &source shape&$R_\sigma$&\ref{eq:source_dist_R}&yes&$[0,\infty[$\\
 \hline
\multirow{5}{*}{\parbox{1.2cm}{spectrum shape}}&regression  slope&$\hat{s}$&\ref{eq:regression}&yes&$[0,\infty[$\\
&regression $r^2$-value&$r^2$&\ref{eq:regression_r2}&no&$[0,1]$\\
&avg. St number&$\overline{\text{St}}$&\ref{eq:spectrum_f_mean}&yes&$[0,\infty[$\\
&std. St number&$\text{St}_\sigma$&\ref{eq:spectrum_f_std}&yes&$[0,\infty[$\\
& $\text{PSD}_\text{max}$ St number & $\text{St}_{L\text{max}}$ & \ref{eq:PSD_max}&yes&$[0,\infty[$\\
\end{tabular}
\end{ruledtabular}
\end{table}
}

\subsubsection{Broadband self-similarity}\label{sec:similarity}
An important property of any aeroacoustic source is the self-similarity or scaling behavior over increasing Mach number. An acoustic spectrum that exhibits self-similarity over the Strouhal number indicates a physical source mechanism that is connected to the flow such as turbulence-induced noise\cite{Lighthill1952}. With the speed of sound $a$ the Strouhal number is defined as
\begin{equation}
    \text{St}=\frac{fD_0}{Ma} \,.
\end{equation}
Spectra collapsing over the use of Helmholtz number 
\begin{equation}
    \text{He}=\frac{fD_0}{a}
\end{equation}
indicate a mechanism connected to acoustic resonances~\citep{Mueller1979} or radiation effects~\citep{Michalke1977}\added[R2C17]{due to spatial coherence of a non-compact source}. If a source is self-similar over one of these frequency types (we refer to the absolute frequency, the Helmholtz number, and the Strouhal number as frequency types), there is a linear dependency between the PSD levels over frequency at different Mach numbers. Thus, we calculate the Pearson correlation coefficients \replaced[R2C05]{$\rho_{ij}$}{$\rho_{jj'}$} between all spectra at different Mach numbers \replaced[R2C05]{$M_i$ and $M_j$}{$M_j$ and $M_{j'}$.} \deleted{where $k$ includes all elements from the upper triangular correlation matrix ($i>j$).}
\replaced[R2C05]{
\begin{equation}
\text{corr}_{k} = \rho_{ij}\big(\text{PSD}(M_i,f),\text{PSD}(M_j,f)\big) \quad\text{for}\quad i>j
\end{equation}}
{
\begin{align}\label{eq:rho_jj}
\rho_{jj'}=&\sum_i^I\big[(\text{PSD}(M_j,f_i))-\langle\text{PSD}(M_j,f_i)\rangle_i)\nonumber\\
&\times(\text{PSD}(M_{j'},f_i))-\langle\text{PSD}(M_{j'},f_i)\rangle_i)\big]\nonumber\\
&\times\big[\sigma_i(\text{PSD}(M_j,f_i)) \sigma_{i}(\text{PSD}(M_{j'},f_i))\big]^{-1}
\end{align}
Note that for the calculation of eq.~\ref{eq:rho_jj} using the Strouhal or Helmholtz number, the spectra have to interpolated on the same Strouhal or Helmholtz number vectors for different Mach numbers. For the results presented in this paper we use linear interpolation on a logarithmic Strouhal and Helmholtz number vector with 12 bins per octave. Since the correlation matrix is symmetric and the diagonal entries are unity, we determine the Mach average $\overline{\mathrm{corr}}$ and standard deviation $\mathrm{corr}_\sigma$ of the correlation coefficients using the upper triangular matrix $j>j'$.
\begin{align}\label{eq:upper_triangular_average}
    \overline{\mathrm{corr}} &= \frac{2}{J(J-2)} \sum \limits_{j > j'} \rho_{jj'} \\
    \mathrm{corr}_\sigma &= \sqrt{\frac{2}{J(J-2)} \sum \limits_{j > j'} (\rho_{jj'} - \overline{\mathrm{corr}})^2}
\end{align}
}

Most aeroacoustic spectra decay in SPL over frequency and if this decay is stronger than local structures or peaks in the spectra, this can result in strong correlations for frequency types over which the spectra are not self-similar. However, the \replaced{correlations $\text{corr}_{k}$}{correlation matrix $\rho_{jj'}$} then often exhibit great variance. Thus, a mean correlation \replaced{$\langle \text{corr}_k \rangle_k$}{$\overline{\mathrm{corr}}$} is not an optimal definition for the self-similarity. It can be improved by taking its standard deviation \replaced{$\sigma_k(\text{corr}_{k} )$}{$\mathrm{corr}_\sigma$} and the mean p-value \replaced{$\langle p_k \rangle_k$}{$\overline{p}$} into account. \added[R2C11]{The p-values are averaged as shown in eq.~\ref{eq:upper_triangular_average} and represent the reliability of the correlation estimation.} Due to the beamforming process in combination with CLEAN-SC, the discrete spectra often contain missing values. If we drop the corresponding frequencies $f$ where \replaced[R2C12]{$\text{PSD}(M,f)=\text{NaN}$}{$\text{PSD}(M,f)\not\in\mathbb{R}$} before the calculation, the standard deviation\deleted{$\sigma$}, and the p-value\deleted{$p$} will increase drastically when the spectra are not self-similar. Using these properties, we introduce the final broadband self-similarity (scal) 
\replaced[R2C05,R2C06]{
\begin{equation}%\label{eq:scalability}
\text{scal} = \big(\langle \text{corr}_{k} \rangle_k - \sigma_k\left(\text{corr}_{k}\right)\big)\big(1-\langle p_k\rangle_k\big)\,,
\end{equation}}{
\begin{equation}\label{eq:scalability}
\mathrm{scal} = (\overline{\mathrm{corr}} - \mathrm{corr}_\sigma)\big(1-\overline{p}\big)
\end{equation}}
which can be calculated over all frequency types separately. This definition strongly penalizes a high p-value and a large variance in the correlations. 
% Figure~\ref{fig:Figure2} $c)$ shows the mean correlation coefficient and its standard deviation, the mean p-value and the resulting self-similarity for a slat tone source over a variety of modification exponents $m$.

\subsubsection{Frequency normalization exponent}\label{sec:mod_exponent}
\begin{figure}
	\centering
	\includegraphics[width=\reprintcolumnwidth]{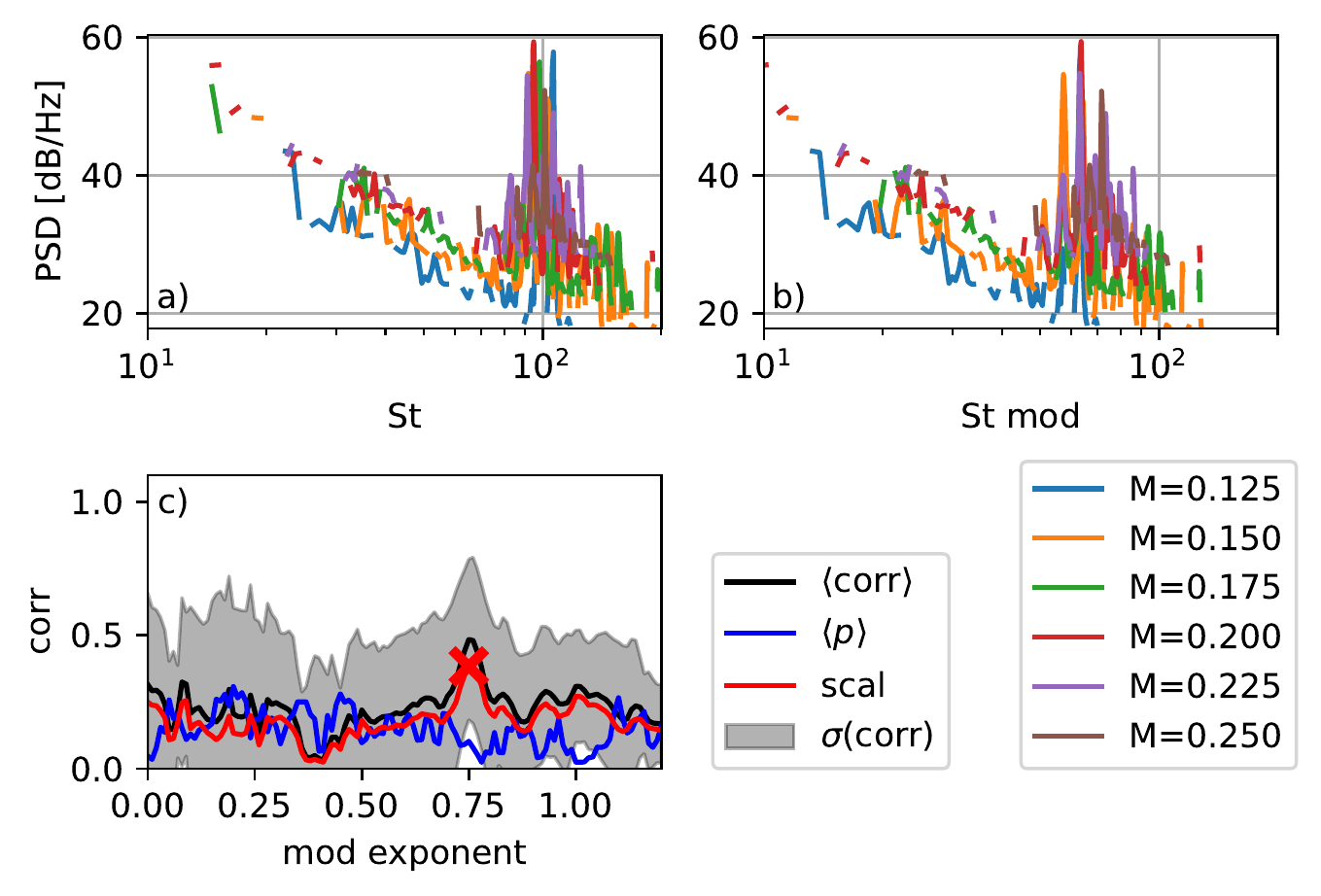}
	\caption{(Color online) Do728, $\text{Re}=\SI{1.4e6}{}$, slat tones at $\alpha=\SI{8}{\degree}$. Comparison of source spectra at different Mach numbers over $a)$ normal Strouhal number and $b)$ modified Strouhal number, see eq.~\ref{eq:mod_exponent}, with $m=0.74$. $c)$ shows the mean and standard deviation $\sigma$ of the Pearson correlation coefficient of the source spectra at different Mach numbers over the variation of the modification exponent, the mean $p$-value, and the resulting self-similarity (scal), see Section~\ref{sec:similarity}. The modification exponent $m=0.74$ achieves the optimal self-similarity, marked with an $x$.}
	\label{fig:Figure2}
\end{figure}

As stated in sub-section~\ref{sec:mod_exponent} a spectrum scales either over the Strouhal or the Helmholtz number. However, we observe in the presented data that \replaced[R1C05, R2C18]{Strouhal-scaling spectra do not perfectly align over the Strouhal number}{spectra that are supposed to scale over the Strouhal number are often not perfectly aligned as depicted in Figure~\ref{fig:Figure1} $a)$}. In this rare case of multiple cryogenic measurement conditions, both datasets allow us to observe spectra at constant \added[R1C05]{absolute} Reynolds number over increasing Mach number (at decreasing temperatures and increasing pressure). \replaced[R1C05]{With the dynamic viscosity $\mu(T)$ and density $\rho(p_0,T)$ of the medium, the Reynolds number is
\begin{equation}\label{eq:Reynoldsnumber}
    \text{Re}=\frac{\rho(p_0,T)M(T,u)D_0}{\mu(T)} \,.
\end{equation}
At a constant Reynolds number, the spectra align perfectly over the Strouhal number at increasing Mach number.}{Using spectra spectra at different Mach numbers from different Reynolds configurations so that the absolute Reynolds number is kept constant (Re(D4, $M$=0.125)=Re(D3, $M$=0.15)=Re(D2, $M$=0.12)=Re(D1, $M$=0.25)=$\SI{2e6}{}$), the spectra are perfectly aligned (not depicted here).} Thus, the increase of Reynolds number over Mach number at constant pressure and temperature may cause sources to decrease or increase the Strouhal numbers dependence on the Mach number. Since cryogenic measurements are rare and expensive, we assume that many datasets are observed at constant pressure and temperature (i.e. different \added[R1C05]{absolute} Reynolds numbers) and are affected by this phenomenon. To overcome this problem we define a modified normalized frequency $\hat{f}$ that \replaced[R2C22]{always allows a collapse of the spectra}{compensates for this altered Mach dependency} by introducing the generalized frequency normalization exponent $m$. We then define the modified normalized frequency as
\begin{equation}\label{eq:mod_exponent}
    \hat{f}=\frac{f D_0}{M^m a} \,.
\end{equation}
Note that this normalized frequency is a generalization of the Helmholtz number (for $m=0$) and the Strouhal number (for $m=1$). For convenience, we will speak of a modified Strouhal number if $m\ge 0.5$. To obtain the generalized frequency normalization exponent, we optimize the collapse of the spectra by maximizing its broadband self-similarity (see Section~\ref{sec:similarity}). Figure~\ref{fig:Figure2} shows the comparison of the $a)$ normal Strouhal number and $b)$ the modified Strouhal number with $m=0.72$. Figure~\ref{fig:Figure2} $c)$ shows the mean spectra correlation \added{$\overline{\text{corr}}$} (black line) its standard deviation \replaced{$1\sigma$}{\added{$\text{corr}_\sigma$}} (gray area) over the modification exponent $m$. The blue line is the mean p-value \replaced{$\langle p\rangle_M$}{$\overline{p}$}, indicating the reliability of the correlation estimation. The optimal value $m^\star$ (shown with the red $x$) is achieved at the global maximum of the self-similarity, see equation~\ref{eq:scalability}. 
\begin{equation}\label{eq:m_star}
    m^{\star} = \underset{m \in [0,\infty[}{\mathrm{argmax}}(\mathrm{scal}(m)) 
\end{equation}

In the example of these slat tones, the increase in Reynolds number results in a weaker Mach dependency of the \added[R2C22]{normalized} frequency than a regular Strouhal number.\\

The following acoustic properties are derived from spectra at different Mach numbers, which can be displayed over the Strouhal number, the Helmholtz number, or the introduced \replaced[R2C23]{generalized}{modified normalized} frequency. \deleted[R2C04f]{There are two possible methods for the calculation of features based on normalized frequency spectra. Either we calculate the features only based on spectra scaled over the generalized frequency, or we calculate the features independently over both the modified Strouhal number and the Helmholtz number. The disadvantage of the first method is that the results depend heavily on the correct determination of the frequency modification exponent. Also, spectra can feature multiple frequency regions with different self-similarities. }Since aeroacoustic experts are used to analyzing spectra displayed over Strouhal and Helmholtz number we\deleted[R2C04f]{ focus on the second approach, i.e. we} calculate features from spectra displayed over both Helmholtz and modified Strouhal number separately. If a spectrum is dominated by a Helmholtz number scaling mechanism, the frequency modification exponent will result in values $m\approx0$. To present spectra over both Helmholtz and (modified) Strouhal number, we have to find a local maximum of the self-similarity function around \replaced{$m\approx1$}{$m_\text{St}\approx1$} to account for spectra that include minor, Strouhal number scaling mechanisms. To do so, we run a standard peak detection over the self-similarity function $\text{scal}(m)$ to find a local maximum \replaced{$m\ge0.5$}{$\tilde{m}\ge0.5$} \added[R2C24]{with a peak prominance $\widehat{P}_\text{p}\ge0.1$}. If none is found and \replaced{$m<0.5$}{$m^\star<0.5$}, we simply set \replaced{$m=1$}{$m_\text{St}=1$} to obtain a (modified) Strouhal number.\\
\added[R2C24]{
\begin{equation}\label{eq:m_tilde}
    \tilde{m} = \underset{\underset{\mathrm{max}(\widehat{P}_\text{p})\ge0.1}{m \in [0.5  ,\infty[}}{\mathrm{peak}}(\mathrm{scal}(m)) 
\end{equation}

\begin{equation}\label{eq:freq_norm_exp}
m_\text{St} = \begin{cases}
    m^{\star} & \mathrm{if} \ m^{\star}>0.5 \\
   \tilde{m}  & \mathrm{if} \ m^{\star}\le0.5  \\
   1 & \mathrm{else}
\end{cases}
\end{equation}
}
\subsubsection{Sound power scaling}\label{sec:Mach_number_scaling}
\begin{figure}
	\centering
	\includegraphics[width=\reprintcolumnwidth]{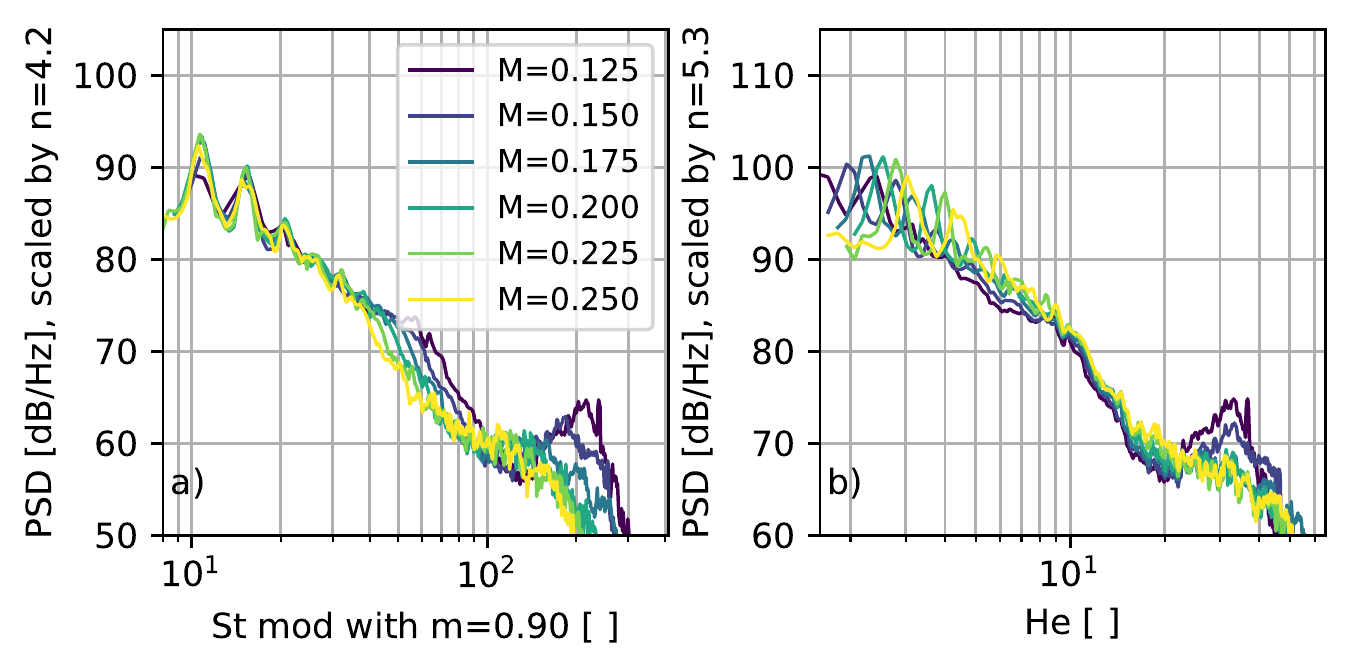}
	\caption{(Color online) Do728 slat / slat track source at $\text{Re}=\SI{1.4e6}{},\alpha=\SI{3}{\degree}$. Comparison of the power scaling over modified Strouhal ($m=0.9$) and Helmholtz number to collapse the PSDs with eq.~\ref{eq:scaled_PSD}. The scaling exponents are $n_\text{St}=4.2$, $n_\text{He}=5.3$, $n_\text{OASPL}=6.3$.}
	\label{fig:Figure3}
\end{figure}
The power of aeroacoustic noise generally increases with increasing Mach number. A prominent example is jet noise for which Lighthill derived the $M^8$ scaling law~\citep{Lighthill1952}. Thus, when doubling the Mach number, the PSD will increase by $10\log_{10}(2^8)=\SI{24}{\decibel}$. The power scaled PSDs ($\widehat{\text{PSD}}$), with the power scaling exponent $n$, are given by
\begin{equation}\label{eq:scaled_PSD}
\widehat{\text{PSD}}(f_i,M_j)=\text{PSD}(f_i,M_j)-n10\log_{10}(M_j)\,.
\end{equation}
Conventionally, a regression on the Overall Sound Pressure Levels (OASPL, which is the sound power integrated over frequency) or peak levels of eq.~\ref{eq:scaled_PSD} is used to determine $n$. This does not always work for spectra from beamforming maps, since the microphone array aperture in combination with deconvolution acts like a low-cut filter at an absolute frequency\deleted[R2C27]{ $f_0$}. This is \deleted{especially }problematic when scaling over the Strouhal number (and Helmholtz number \deleted{when scaling spectra }at different temperatures and pressures)\replaced[R1C06]{ because this results in a Mach number dependent low-cut frequency and thus, in a wrong scaling exponent. To avoid this problem,}{. It effectively creates a Mach dependent low-cut filter which in combination with a typical SPL decrease over frequency for aeroacoustic sources results in a wrong OASPL and thus, scaling exponent. Instead,} we minimize the mean distance between all power scaled spectra $\widehat{\text{PSD}}(f_i,M_j)$ over frequency bin-wise with a standard bounded minimization algorithm. To calculate a distance between multiple spectra at once we use the standard deviation $\sigma$ of the spectra at every discrete frequency. Since parts of a spectrum with a high SPL are often considered more important for the scaling, we can \replaced[R2C28]{regularize}{weight} the standard deviations at every frequency with the Mach averaged spectrum power \replaced{$\langle \text{PSD}(f_i,M_j)\rangle_{M_j}^\gamma$ with $\gamma\ge0$}{$\langle \text{PSD}(f_i,M_j)\rangle_{M_j}^\kappa$}. The hyperparameter \replaced{$\gamma$}{$\kappa$} of this \replaced[R2C28]{regularization}{weight} determines by how much we want to prefer the scaling of high levels. Thus, we minimize
\replaced[R1C27]{
\begin{multline}%\label{eq:Mach_scaling}
\min\limits_{0\le n<\infty} \sum_{i}\bigg[\sigma_{j} \big(\text{PSD}(f_i,M_j)-n10\log_{10}(M_j)\big) \\
\big\langle \text{PSD}(f_i,M_j)\big\rangle_{j}^\gamma\bigg]\,,
\end{multline}
}{\begin{multline}\label{eq:Mach_scaling}
\min\limits_{0\le n<\infty} \sum_{i}\bigg[\sigma_{j} \big(\text{PSD}(f_i,M_j)-n10\log_{10}(M_j)\big) \\
\big\langle \text{PSD}(f_i,M_j)\big\rangle_{j}^\kappa\bigg]\,,
\end{multline}}
with \replaced[R1C27]{$n,\gamma\in\mathbb{R}\ge0$}{$n,\kappa\in\mathbb{R}\ge0$}. For the calculation of a reliable power scaling exponent at least spectra at three different Mach numbers should be used. A large variation in Mach number also increases the scaling's reliability. Figure~\ref{fig:Figure3} shows the resulting power scaling for a slat / slat track source and $\gamma=10$ over modified Strouhal number and Helmholtz number. Note the different scaling behavior over Strouhal and Helmholtz number for the low and high-frequency part of the spectrum and that the OASPL scaling \added[R2C26]{$n_\text{OASPL}$ obtained by a linear regression} neither matches the Helmholtz nor the Strouhal scaling exponent correctly, as described above. In Figure~\ref{fig:Figure7} more examples of scaled PSDs with $\gamma=10$, which was used for all \replaced{sources}{results} presented in this paper, are displayed.

\subsubsection{Tonality}\label{sec:tonality}
\begin{figure}
	\centering
	\includegraphics[width=\reprintcolumnwidth]{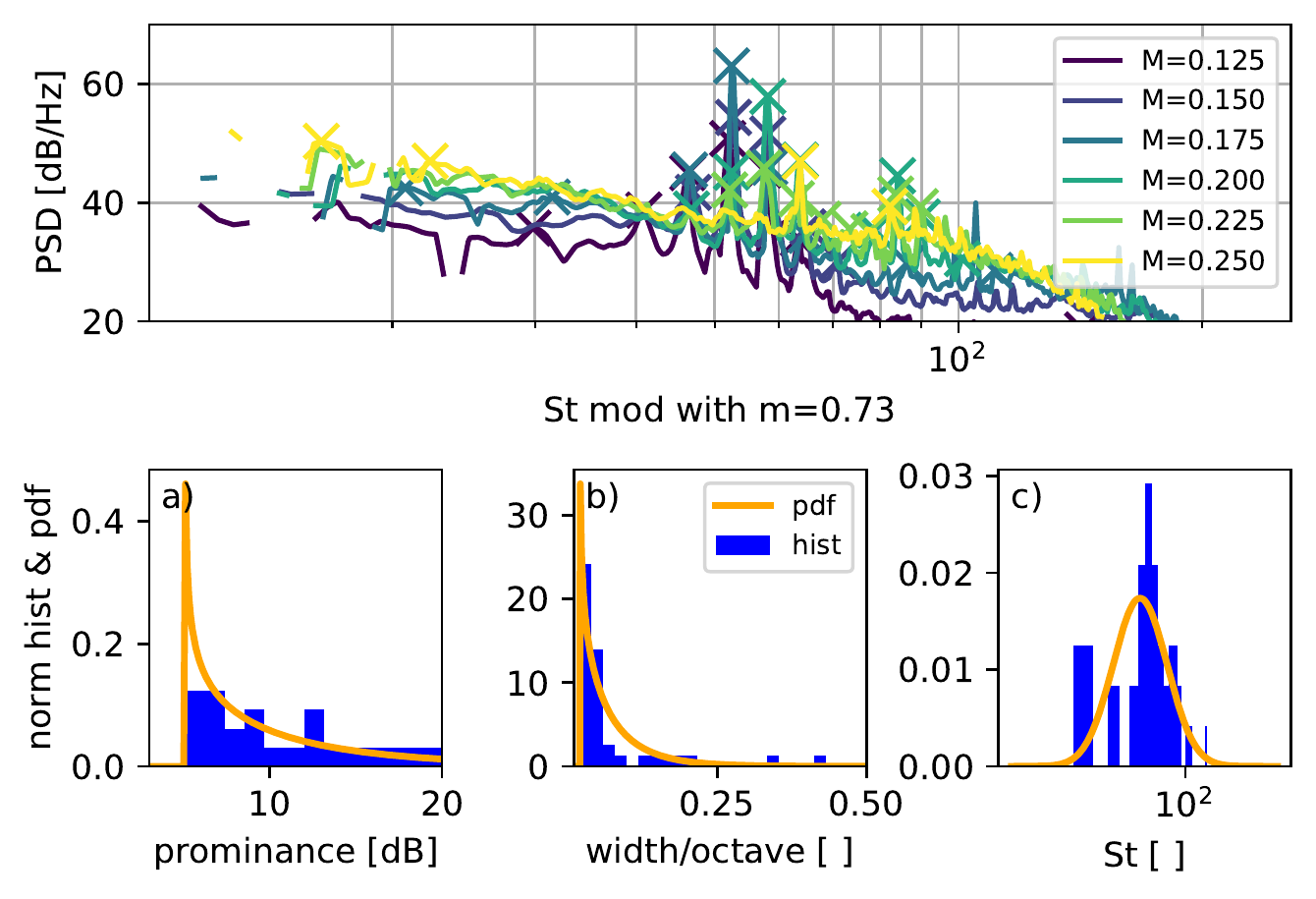}
	\caption{(Color online) Do728, $\text{Re}=\SI{1.4e6}{}$, slat tones at  $\alpha=\SI{8}{\degree}$. Top, the spectra and the automatically detected peaks, depicted with an $x$. Bottom, a normalized histogram and the fitted PDF of the $a)$ peaks prominence, $b)$ peak width, and $c)$ peak Strouhal number.}
	\label{fig:Figure4}
\end{figure}
Accounting for the tonal behavior of the sources \deleted[R1C30]{is a complex task and }results in less straightforward feature descriptions, since the \deleted[R1C30]{number of} tonal peaks \deleted{$P$} vary from source to source and within a source for different Mach numbers but \replaced[R1C30]{we have to define a constant amount of feature values that describe their properties.}{their properties must be captured by a finite amount of features, see section~\ref{sec:features}.} These properties are the peak-width intervals \replaced{$P_w$}{$\widehat{P}_\text{w}$}, peak prominences \replaced{$P_p$}{$\widehat{P}_\text{p}$}, peak frequencies \replaced{$P_f$}{$\widehat{P}_\text{f}$}, and the number of peaks \replaced{$P_n$}{$\widehat{P}_\text{n}$}. \deleted[R1C30]{We also want to know whether the peaks scale well over the modified Strouhal or Helmholtz number ($\text{scal}_\text{p}$). }First, we run a standard automated peak detection over the spectra\added[R1C30]{, which results in a set of peak prominences $P_\text{p}(M_j)$, peak widths $P_\text{w}(M_j)$, and peak frequencies $P_\text{f}(M_j)$ for every Mach number $M_j$. Additionally, we define sets of frequency bins $P_{\tilde{\text{f}}}(M_j)$ that include all frequency bins, that lie within the peak width intervals $P_\text{w}(M_j)$. 
\begin{equation}
     P_{\tilde{\text{f}}}(M_j) = \lbrace f_i | s.t. \ f_i  \ \text{belongs to a peak for}\ M_j    \rbrace    
\end{equation}
Thus, the number of elements $|\dots|$ of the sets are $|P_\text{p}(M_j)|=|P_\text{w}(M_Fj)|=|P_\text{f}(M_j)|\le|P_{\tilde{\text{f}}}(M_j) |$}. The number of peaks \replaced{$P_n$}{$\widehat{P}_\text{n}$} is \deleted[R1C30]{a straightforward value, and we take their average number from the spectra at different Mach numbers.}{defined as the Mach averaged number of elements in the sets.} 
\replaced[R1C30]{
\begin{equation}%\label{eq:tonality_number}
\widehat{P}_n=\langle P_n(M_j)\rangle_{j}
\end{equation}
}{\begin{equation}\label{eq:tonality_number}
\widehat{P}_\text{n}=\langle |P_p(M_j)|\rangle_{j}
\end{equation}}
As stated in section~\ref{sec:features}, we have to break down the properties of the peaks to single value features. We do so by describing the distribution of the peaks instead of using the individual peaks' properties directly. Naturally, peaks with lower prominence appear more often than peaks with very high prominence \added[R1C30]{in $P_\text{p}$}. This assumption is also used for the peak width \added[R1C30]{in $P_\text{w}$}. This behavior can be modeled using a gamma distribution~\citep{Abramowitz1974}. With the gamma function $\Gamma(k)$, the Probability Density Function (PDF)  $\gamma$ of the gamma distribution over \replaced{$x$}{a variable $v$} with a shape $k$, scale $\theta$ and location $l$ is
\replaced[R1C30]{
\begin{equation}
\gamma(x,k,\theta,l)=\frac{(x-l)^{k-1}\exp(-\frac{x-l}{\theta})}{\theta^k\Gamma(k)} \quad \text{for} \quad x,k,\theta\ge0\,.
\end{equation}
}
{\begin{equation}\label{eq:tonality_gamma}
\gamma(v,k,\theta,l)=\frac{(v-l)^{k-1}\exp(-\frac{v-l}{\theta})}{\theta^k\Gamma(k)} \quad \text{for} \quad v,k,\theta,l\ge0\,.
\end{equation}
}

For the peak frequency locations \added{in $P_\text{f}$}, we found that a log-normal distribution fits most of the sources best. The distributions of these properties have unknown shape (standard deviation), scale (distribution median), and location (distribution offset) parameters which can be approximated from any number of samples greater than one by fitting the gamma or log-normal distribution to the data with standard fitting methods. The \replaced[R1C28]{slope}{shape}, scale, and location for \deleted[R1C29]{each} peak width, peak prominence, and peak frequency are used as comparable feature values, independent of the number of tones in the PSDs and the number of different Mach numbers. Figure~\ref{fig:Figure4} shows the peak detection and the corresponding distributions for all Mach numbers of a Do728 slat tone source. We set a lower threshold of $\SI{3}{\decibel}$ for the peak prominence to prevent the algorithm from detecting lots of irrelevant low-level peaks which dominate the distributions. We set the feature values to zero if only one or fewer tones are detected.\\

To determine how well the prominent peaks scale over Strouhal or Helmholtz number \added[R2C33]{($\text{scal}_\text{p}$)} we average the ratio of how many peaks overlap at every detected peak frequency interval. Working with logarithmically spaced, discrete frequencies $f_i$, we introduce two sets. \replaced[R2C33]{$A_i$ are the sets of Mach numbers that feature a peak $P_w(M_j)$ at the corresponding frequency $f_i$. $B$ is the single set of unique frequencies $f_i$ for which at least one spectrum $\text{PSD}(M_j)$ features a peak $P_w(M_j)$.}{$E_i$ are $I$ sets, each containing up to $J$ Mach numbers for which the frequency bin of the corresponding spectrum lies within a peak interval. $\widehat{I}$ is a single set, that contains the the indices $i$ for which there exists at least one Mach number for which the frequency lies within a peak interval.}
\replaced[R2C33]{
\begin{align}
A_i &=& \{ M_j |             \quad &\text{so that}\quad f_i \in P_w(M_j)\}\\
B &=& \{  f_i | \exists M_j \quad &\text{so that} \quad f_i \in P_w(M_j)\} 
\end{align}}{
\begin{align}
E_i &=& \{ M_j |             \quad &\text{so that}\quad f_i \in P_{\tilde{\text{f}}}(M_j)\}\\
\widehat{I} &=& \{  i | \exists M_j \quad &\text{so that} \quad f_i \in P_{\tilde{\text{f}}}(M_j)\} 
\end{align}}
The resulting scaling of the tones is then the ratio of spectra that share a peak at the same frequency to the total number of spectra at different Mach numbers \replaced{$|M_j|$}{$J$}, averaged over all frequency bins $i$ for which at least one peak was detected
% \todo[inline]{$|M_j|$ ist der Betrag der j-ten Mach-Zahl also $M_j$ selbst. Für die Anzahl der Machzahlen hatten wir oben schon $J$ benutzt $\rightarrow$ Muss aber ganz am Anfang des Kapitels noch eingeführt werden.}
\replaced{
\begin{equation}%\label{eq:tonality_scal_p}
\text{scal}_{p}=\bigg\langle\frac{|A_i|-1}{|M_j|-1}\bigg\rangle_{i}\,.
\end{equation}}
{
\begin{equation}\label{eq:tonality_scal_p}
\text{scal}_\text{p}=\frac{1}{|\widehat{I}|} \sum \limits_{i\in \widehat{I}} \frac{|E_i|-1}{J-1} \,.
\end{equation}}
The minus ones ensure a soft feature value $0\le \text{scal}_\text{p}\le1$ for the modified Strouhal and Helmholtz number, since each $E_i$ contains at least one element. \deleted{Again, if $A$ is empty (no peaks exist, thus, they do not scale) we set $\text{scal}_p=0$.} Finally, we introduce the tonal intensity $\text{prop}_\text{p}$, defined as the Mach averaged ratio of tonal SPL to total SPL. It expresses how much percent of the energy in the spectra is caused by tones.
\replaced{
\begin{equation}%\label{eq:tonality_prop_p}
\text{prop}_p=\left \langle\frac{\sum_{i} \text{PSD}(f\in B,M_j)}{\sum_j\text{PSD}(f_i,M_j)}\right \rangle_{j}
\end{equation}
}{
\begin{equation}\label{eq:tonality_prop_p}
\text{prop}_\text{p}=\left \langle\frac{\sum_{f_i \in  P_\text{w}(M_j)} \text{PSD}(f_i,M_j)}{\sum_i\text{PSD}(f_i,M_j)}\right \rangle_{j}
\end{equation}
}
\subsubsection{Source location dependency on the Mach number}\label{sec:srs_mov}
\begin{figure}
	\centering
	\includegraphics[width=\reprintcolumnwidth]{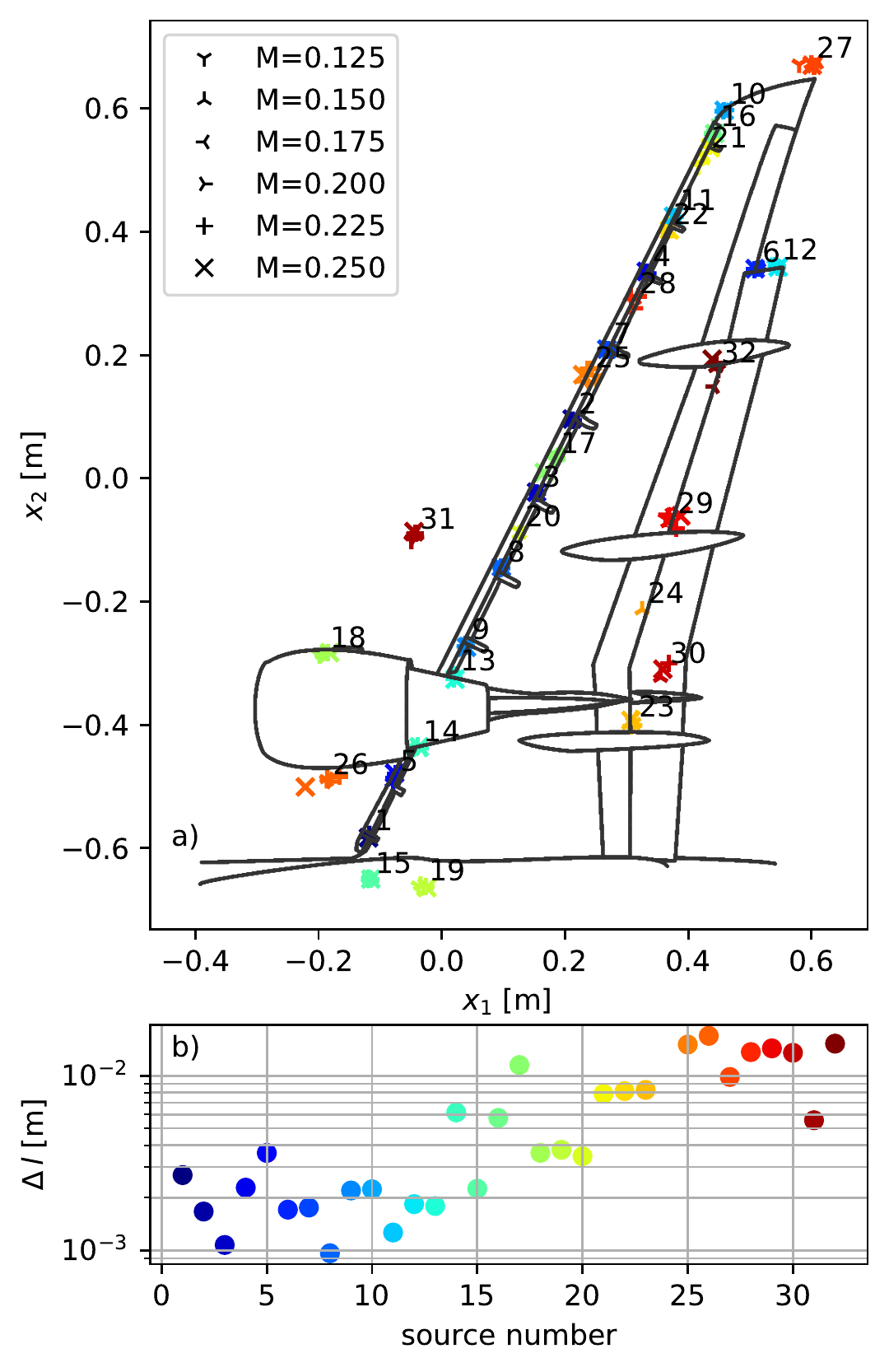}
	\caption{(Color online) Do728. $a)$ shows the source positions at $\text{Re}=\SI{1.4e6}{}, \alpha=\SI{10}{\degree}$ (if the source is present at this configuration) for the given Mach numbers. $b)$ shows the mean movement of the sources calculated with eq.~\ref{eq:srs_mov}.}
	\label{fig:Figure5}
\end{figure}
The spatial location of some aeroacoustic sources may change with the Mach number. An example of a moving source would be a flow detachment, at which the sound generating eddies move further downstream with increasing Mach number or jet noise, while cavity noise would remain at the same location. Figure~\ref{fig:Figure5}\replaced{, top,}{ $a)$} shows the variation of source positions on the Do728 with increasing Mach number. For a numeric feature description, we calculate the positional change of the source with increasing Mach number and call it the source movement. We define the source movement $\Delta l$ as the mean movement of the local source position $\vec{x}$, normalized by the change in Mach number, shown in Figure~\ref{fig:Figure5}\replaced{, bottom}{ $b)$}. A limitation of this feature is that it assumes the monotonous movement of the source in one direction with the Mach number.
\begin{equation}\label{eq:srs_mov}
\Delta l = \left\langle{ \frac{|\vec{x}_{j}-\vec{x}_{j-1}|}{M_{j}-M_{j-1}}}\right\rangle_{j}\quad \text{for}\,j\ge2
\end{equation}

\subsubsection{Spatial source distribution}\label{sec:srs_distribution}
Aeroacoustic sources can be spatially distributed, such as line or volume sources. We use the sources' spatial PDFs obtained with SIND to describe this behavior. SIND approximates the spatial source distributions with 2D normal distributions, described by the standard deviations $\sigma_{x_i}$. We define the integrated, normalized PDF area $A$ as a feature for the \replaced[R2C36]{spatial expansion of the source distribution.}{spatial compactness of the source.}  
\replaced[R1C07, R2C34]{
\begin{equation}
A = \int_{x_1}\int_{x_2}\text{PDF}(x_1,x_2)\mathrm{d}x_2\mathrm{d}x_1
\end{equation}
}
{
\begin{align}\label{eq:source_dist_A}
    A & = \int_{x_1}\int_{x_2}\frac{\text{PDF}(x_1,x_2)}{||\text{PDF}(x_1,x_2)||_\infty}\mathrm{d}x_2\mathrm{d}x_1\nonumber\\
      & = 2 \pi \sigma_{x_1} \sigma_{x_2}
\end{align}
}
% \todo[inline]{\[ A = 2 \pi \sigma_{x_1} \sigma_{x_2}  \]}
We define the ratio of the PDFs standard deviations $\sigma_{x_i}$ as an indicator for line sources with
\begin{equation}\label{eq:source_dist_R}
R_\sigma = \text{max}\left(\frac{\sigma_{x_1}}{\sigma_{x_2}},\frac{\sigma_{x_2}}{\sigma_{x_1}}\right)-1 \,.
\end{equation}
Thus, \replaced{$R\approx1$}{$R_\sigma\approx0$} indicates a point or sphere-like source while an increasing \replaced{$R$}{$R_\sigma$} indicates a line source.

\subsubsection{Spectrum shape}\label{sec:spectrum_shape}
To capture the general spectrum shape, we use a linear regression $L(f)$ for $\text{PSD}(f)$ which consists of two values: the interception \replaced{$b$}{$i_0$} of the line at $f=\SI{0}{Hz}$ and the slope \replaced{$a$}{$\hat{s}$}
\replaced{
\begin{equation}
L(f) = af+b\,.
\end{equation}
}
{\begin{equation}\label{eq:regression}
L(f) = \hat{s}f+i_0\,.
\end{equation}}
The interception is an absolute value, varies with the Mach number, and is, therefore, discarded. The slope is the increase or decrease of the PSD level over the frequency. Additionally, we use the regression's $r^2$-value which describes how well the linear regression explains the spectrum. A low $r^2$ value indicates that the linear regression is not capturing the movement in the spectrum well. Thus, it is an indication of the waviness of the spectra. \added{With the error $e$ of the regression model
\begin{equation}
e_{ij} = \text{PSD}(f_i,M_j)-L(f_i,M_j)
\end{equation}
the Mach-averaged $r^2$ value is then calculated with 
\begin{equation}\label{eq:regression_r2}
r^2 = \left\langle 1-\frac{\sum_i e_{ij}}{\sum_i (\text{PSD}_{ij}-\langle\text{PSD}_{ij}\rangle_i)^2} \right\rangle_j
\end{equation}
}

Similar to the spatial source distribution, we define a source distribution over frequency. Since we work on sparse spectra which are not defined on all frequency bins, we use the mean \replaced{$f_\text{mean}$}{$\overline{f}$} and standard deviation \replaced{$f_\text{std}$}{$f_\sigma$} of the \replaced{existing frequencies to capture in which frequency interval the source exists}{frequencies for which the source PSD is real valued to capture the source's radiation frequency interval}. With the sets \replaced[R2C35]{$C_j$}{$Q_j$} that contain the \replaced[R2C12]{defined}{real valued} frequency bins $f_i$ for the spectra at Mach number $M_j$
\replaced[R2C35]{
\begin{equation}
C_{ij} = \{ f_i|\quad \text{so that}\quad \text{PSD}(M_j,f_i)\neq \text{NaN} \}\\
\end{equation}
the
\begin{align}
f_\text{mean} & = & \langle \langle \text{PSD}(M_j,f_i)\rangle_i \rangle_{j}\\
f_\text{std} & = & \langle \sigma_{i} (\text{PSD}(M_j,f_i)) \rangle_j
\end{align}
}
{
\begin{align}
Q_j &= \{ f_i| \quad \text{such that} \quad \text{PSD}(f_i,M_j) \in\mathbb{R} \} \\
\overline{f} &= \frac{1}{J} \sum_j\left(  \frac{1}{|Q_j|}\sum_{f \in Q_j} f\right) \label{eq:spectrum_f_mean}\\
f_\sigma &=  \sqrt{ \frac{1}{J} \sum_j  \left(\frac{1}{|Q_j|}\sum_{f \in Q_j} (f - \overline{f})^2\right) \label{eq:spectrum_f_std}} 
\end{align}
}

\added{Finally, we introduce the frequency $f_{L\text{max}}$ for which the PSD level has a maximum.
\begin{equation}\label{eq:PSD_max}
f_{L\text{max}} = \langle \text{argmax}_i(\text{PSD}(f_i,M_j))\rangle_j
\end{equation}

}
\subsection{Source clustering}\label{sec:clustering}
\begin{figure}
	\centering
	\includegraphics[width=\reprintcolumnwidth]{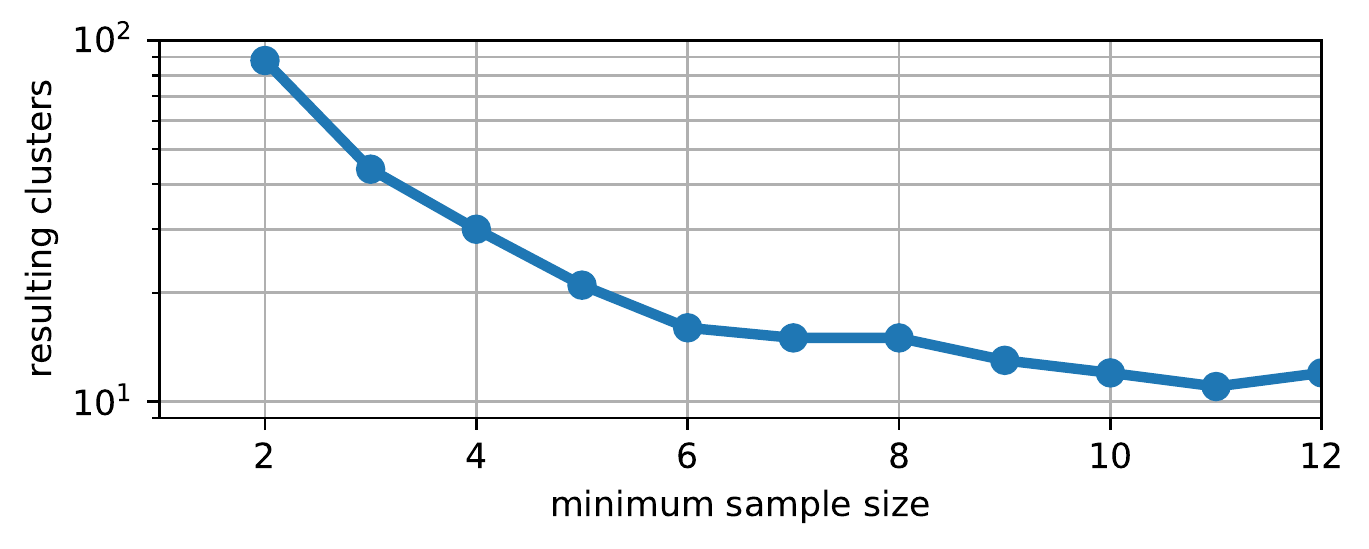}
	\caption{(Color online) Do728, the figure shows the resulting numbers of clusters from HDBSCAN over the minimum sample size parameter.}
	\label{fig:Figure6}
\end{figure}

We employ unsupervised clustering that groups the sources based on \replaced{their}{a} distance \added{metric} in the introduced aeroacoustic feature space.\deleted[R1C31]{The features do not include the Reynolds numbers the angle of attack, and the absolute or relative source positions.} We select Hierarchical Density-Based Spatial Clustering of Applications with Noise (HDBSCAN)~\citep{Campello2013,McInnes2017}, which supports soft clustering without prior knowledge about the number of clusters, i.e., an expectation of the number of different source types. HDBSCAN also provides a clustering confidence. Since the distances between sources in a high dimensional feature space become alike~\citep{Aggarwal2001}, the feature space must be reduced. The most prominent dimensionality reduction technique is Principal Component Analysis, which orthogonalizes the feature space and sorts the dimensions based on their explained variance. Then, dimensions with little statistical variance can be discarded. For the presented method, we use a Kernel Principal Component Analysis (KPCA)~\citep{KPCA} with a Radial Basis Function (RBF) as a kernel. The KPCA uses a nonlinear convolution kernel, which allows the PCA to embed the feature space in a nonlinear manifold. For dimensionality reduction, we retain $2\sigma\approx\SI{95}{\percent}$ of the explained variance by discarding the KPCA dimensions with the least variance. Before using a PCA or KPCA, the feature space must be normalized to zero mean and unity variance. This normalization assumes that the features are normally distributed. However, most of the features $F$ are distributed exponentially and are transformed to a log-space prior to the KPCA with $\log_{10}(F+1)$, \replaced{i.e. all features but the source movement, the generalized frequency exponent, the source compactness, the broadband and tonal scaling behavior, the Mach power scaling, and the linear regression's $r^2$ value}{see table~\ref{tab:features_table}}. Thus, the feature space is partially transformed into a log-space, then normalized, then orthogonalized and reduced to $\SI{95}{\percent}$ explained variance using a KPCA, and then clustered using HDBSCAN.\\

HDBSCAN requires a minimum sample size which determines how many source members a cluster must have. Smaller clusters are discarded as noise. We can determine a reasonable sample size by examining the resulting cluster numbers. With an increasing sample size, the resulting number of clusters first drops massively and then decreases only slightly. Figure~\ref{fig:Figure6} demonstrates this for the Do728 dataset. We pick the sample size after which the total quantity of clusters $n_\text{clusters}$ only decreases slightly, in this case around $n_\text{samples}=7$. This is sometimes referred to as the ``elbow-method'' or ``knee-method''~\citep{Thorndike53}. It is noteworthy that picking a bad sample size may result in sub-clusters (splitting a cluster) or super-clusters (merging clusters), but the overall results remain comparable. Thus, it can be helpful to start with a large sample size to obtain few clusters which are manageable to analyze manually and then transition towards smaller sample sizes. We call the described methodology for this EDSS ``\textbf{C}luste\textbf{r}ing sources based on their \textbf{a}eroacoustic \textbf{f}ea\textbf{t}ures'' (CRAFT).

\section{Manual source identification}\label{sec:classification}
\begin{figure}
	\centering
	\includegraphics[width=\reprintcolumnwidth]{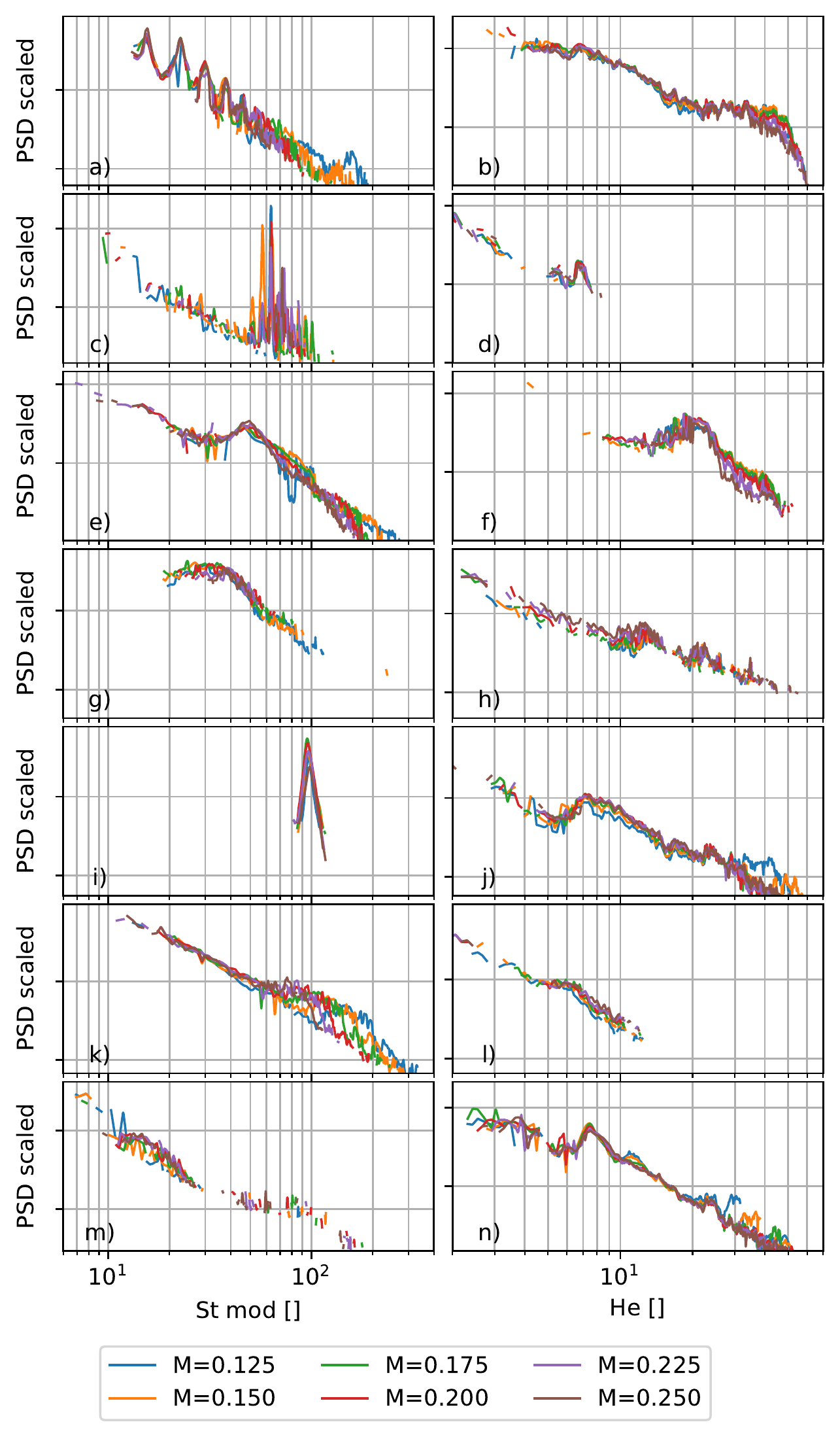}
	\caption{(Color online) Do728, normalized spectra of the typical source types: $a)$ slat, $b)$ slat track, $c)$ slat tone, $d)$ flap tone, $e)$ trailing flap side edge, $f)$ leading flap side edge, $g)$ strake at low Reynolds number and high angle of attack, $h)$ wind tunnel, $i)$ strake tone at high Reynolds number and low angle of attack, $j)$ outer nacelle, $k)$ slat edge, $l)$ flap track, $m)$ wing tip, and $n)$ slat resonance. The horizontal grid lines depict $\Delta\SI{20}{\decibel}$. The frequency modification exponents $m$, the Mach scaling exponents $n$, and the corresponding angles of attack $\alpha$ and Reynolds numbers are given in Table~\ref{tab:sourcetype_parameters}.}
	\label{fig:Figure7}
\end{figure}

\added[R1C08]{
\begin{table}%[ht!]
\caption{\label{tab:sourcetype_parameters}Generalized frequency exponents $m$, Mach scaling exponents $n$, angles of attack $\alpha$, and Mach averaged Reynolds numbers $\langle \text{Re}\rangle_M$ for the displayed sources in Fig.~\ref{fig:Figure7}.}
\begin{ruledtabular}
\begin{tabular}{r|ccccccc}
 & $a)$& $b)$&$c)$&$d)$&$e)$&$f)$&$g)$ \\ 
\hline
$m$& 0.90& 0.00& 0.75& 0.00& 1.09& 0.00& 0.71\\
$n$& 4.44& 5.47& 3.43& 4.23& 4.43& 4.09& 3.43\\
$\alpha$ [$\si{\degree}$] & 3& 5& 9& 9& 3& 5& 9\\
$\langle \text{Re}\rangle_M$ [\SI{1e6}{}] & 1.4 & 1.4 & 1.4 & 1.4 & 1.4 & 10.6 & 1.4\\
\hline\hline
&$h)$&$i)$&$j)$&$k)$&$l)$&$m)$&$n)$\\
\hline
$m$& 0.00& 0.94& 0.00& 1.00 & 0.00& 0.66& 0.00\\
$n$& 3.87& 3.68& 5.47& 3.26& 6.07& 3.59 & 5.26\\
$\alpha$ [$\si{\degree}$] & 10& 3& 5& 10& 3& 9& 1\\
$\langle \text{Re}\rangle_M$ [\SI{1e6}{}] & 2.5& 10.6& 1.4& 10.6& 1.4& 1.8& 1.4 
\end{tabular}
\end{ruledtabular}
\end{table}
}

To visualize, analyze, and quantify the results of the EDSS, we first manually identify source types and label the sources in the Do728 and A320 datasets based on their spectra and their self-similarity. The labels for the manually identified source types are mainly chosen based on the sources' spatial \replaced{appearance}{location}. We also compare the source spectra to each other to identify sub and super-categories. To make this process transparent to the reader, we present in Figure~\ref{fig:Figure7} exemplary Do728 spectra for the most common categories. The sources in the left column are displayed over the modified Strouhal number and sources in the right column are displayed over the Helmholtz number. We note that the manual source type identification and label choices may be ambiguous, contain errors, and misinterpretations. Also, we emphasize that the source groups and corresponding labels cannot be held as ground truth, since they have not been obtained by independent researchers.

\begin{itemize}
\itemsep-0.1em 
\item The slats feature Strouhal number scaling peaks with overtones that decay in level and prominence with increasing frequency, see Figure~\ref{fig:Figure7} $a$). They are mainly located at or between the slat tracks, see Figure~\ref{fig:Figure5} source locations 2, 4, 8, 9, 17, 20, 21, and 28.
\item The slat tracks scale over Helmholtz number, see Figure~\ref{fig:Figure7} $b$). At high frequencies they often exhibit a Helmholtz scaling hump that is Mach number dependent. In Figure~\ref{fig:Figure5} they are at the source locations 2, 3, 4, 7, 8, 9, 11, and 16.
\item The slat tones feature extremely dominant Strouhal number scaling small-band tones, see Figure~\ref{fig:Figure7} $c$). They are mainly located at the slat positions, see Figure~\ref{fig:Figure5} locations 4, 7, 9, 11, 16, 21, 22, 25, and 28.
\item The flap (track) tones feature a small Helmholtz scaling tone and are a sub-category of the flap track, see Figure~\ref{fig:Figure7} $d$). In Figure~\ref{fig:Figure5} they are at the locations 23, 29, and 32.
\item The trailing flap side edges (TFSE), Figure~\ref{fig:Figure5} at location 12, feature a prominent Strouhal scaling peak, see Figure~\ref{fig:Figure7} $e$). 
\item The leading flap side edges (LFSE), Figure~\ref{fig:Figure5} at location 6, feature a smaller Helmholtz scaling peak, see Figure~\ref{fig:Figure7} $f$). At increasing Reynolds numbers, a second Helmholtz number scaling peak emerges. 
\item The strakes feature a Strouhal number scaling hump, see Figure~\ref{fig:Figure7} $g$). It increases in intensity with increasing angle of attack and disappears with increasing Reynolds number. In Figure~\ref{fig:Figure5} they are at the locations 18, and 26.
\item The wind tunnel noise, Figure~\ref{fig:Figure5} at location 31, scales over Helmholtz number, see Figure~\ref{fig:Figure7} $h)$. It appears next to the wing and is considered as a spurious noise source in this measurement~\citep{Ahlefeldt2013}.
\item The strake tone, Figure~\ref{fig:Figure5} at location 18, is a dominant Strouhal scaling tone, see Figure~\ref{fig:Figure7} $i$). It appears only at high Reynolds numbers and low angles of attack and decreases in intensity with increasing angle of attack. It is a sub-category of the strake.
\item The outer nacelle area, Figure~\ref{fig:Figure5} source location 13, features a broadband hump that scales over Helmholtz number, see Figure~\ref{fig:Figure7} $j$).
\item The slat edge is located close to the wing tip, Figure~\ref{fig:Figure5} position 10, and its noise scales over Strouhal number, see Figure~\ref{fig:Figure7} $k$). The spectrum level decays over frequency and features an additional, low-level Helmholtz scaling hump at high frequencies. It is a sub-category of the slat track.
\item The flap tracks feature a low-level Helmholtz number scaling hump, see Figure~\ref{fig:Figure7} $l$). They are located in Figure~\ref{fig:Figure5} at locations 23, 29, and 32.
\item The wing tip, Figure~\ref{fig:Figure5} source location 27, features a Strouhal number scaling hump, see Figure~\ref{fig:Figure7} $l$), that increases in intensity with increasing angle of attack. The spectra are often contaminated with wind tunnel noise.
\item The slat (track) resonances exhibits strong, Helmholtz number scaling peaks and are a sub-category of the slat track, see Figure~\ref{fig:Figure7} $n$). In Figure~\ref{fig:Figure5} they are at the slat tracks 1, and 5.
\end{itemize}

The fuselage (Figure~\ref{fig:Figure5} location 19), nacelle track (appearing only at very high Reynolds numbers at location 30 in Figure~\ref{fig:Figure5}, scaling over Helmholtz number with a very high Mach power exponent $n\approx7.02$), inner slat gap (Helmholtz scaling hump, similar to the LFSE, location 15 in Figure~\ref{fig:Figure5}), and flap gap (location 24 in Figure~\ref{fig:Figure5}) are identified and named based on their spatial appearance. Noise occurring at the flap gap was caused by loose tape on the model during the measurement. Additionally, the category slat / slat track is introduced to account for various spectra that are located on the slat or slat tracks but are ambiguous, e.g. the slat / slat track shown in Figure~\ref{fig:Figure3}, containing some Strouhal number scaling low-frequency peaks, and some Helmholtz number scaling high frequencies.\\
\begin{figure}
	\centering
	\includegraphics[width=\reprintcolumnwidth]{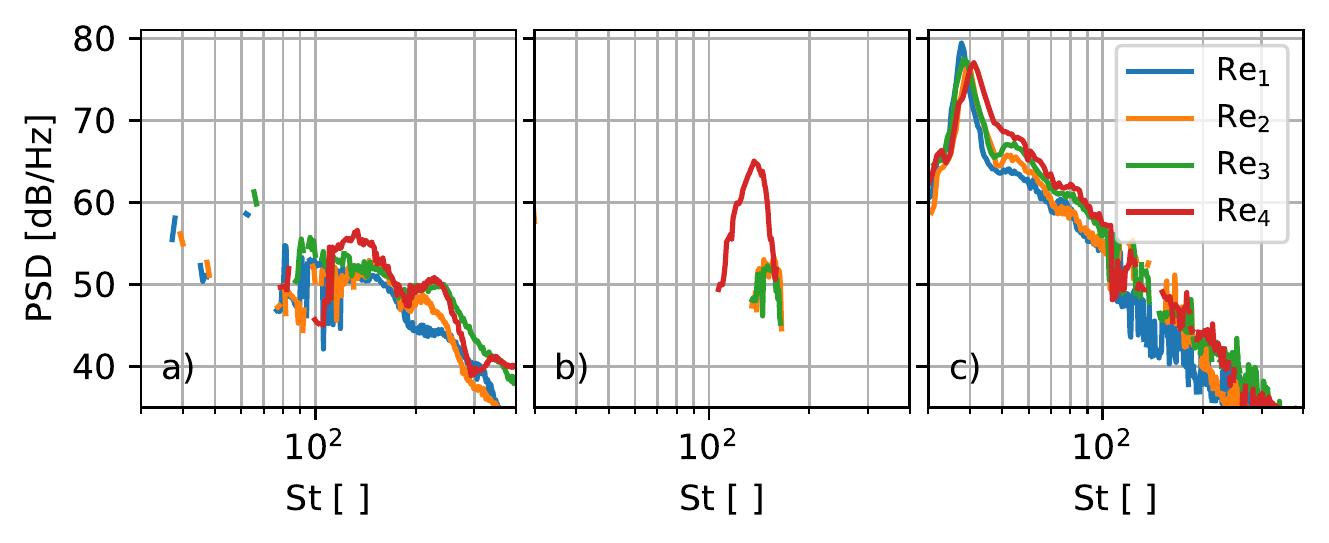}
	\caption{(Color online) A320. The figure shows the Reynolds number effect on the source types $a)$ leading flap side edge at $\alpha=\SI{7}{\degree}$, $M=0.175$, $b)$ high Strouhal number scaling tone at $\alpha=\SI{9}{\degree}$, $M=0.200$, located on the slat and flap, and $c)$ trailing flap side edge at $\alpha=\SI{7}{\degree}$, $M=0.200$.}
	\label{fig:Figure8}
\end{figure}

The manual source type identification in the A320 dataset is more challenging. The smaller microphone array results in less reliable spectra, especially at low frequencies, and due to the small variation in Mach number the correct scaling behavior and scaling exponent are difficult to identify. Additionally, the typical spectra do not consistently correlate with the spatial appearance of the sources, e.g. the spectra are different for different flap tracks. Finally, the sources are strongly affected by the large range of Reynolds numbers which often result in sources that are transitioning from one mechanism to another. Thus, there are multiple sources that we assign the same label, but that feature different spectra (e.g., the fuselage, the slat tracks, and leading flap side edge, see Figure~\ref{fig:Figure7} and vice versa. However, many of the sources are similar to the one found in the Do728 dataset (e.g., LFSE, TFSE, strake, strake tone, slat, slat resonance, slat tone). In addition to the Do728 source types, there exist multiple A320 sources on the slat and flap that feature a high Strouhal number scaling tone (high St tone) that increases in intensity with increasing Reynolds number like the strake tone, see Figure~\ref{fig:Figure8} $b)$. Also, with increasing Reynolds number the peak frequency of the slat, the slat tone, the strake, and the strake tone decreases (with a St mod exponent of $m\approx0.7$) while it increases for the trailing flap side edge ($m\approx1.1$), see Figure~\ref{fig:Figure8} $c)$.\\

\section{Results}
The result section is separated into two parts. First, the results of the aeroacoustic properties and features are presented. Then, based on the resulting feature space, the clustering results are presented.

\subsection{Aeroacoustic feature results}

\begin{figure}
	\centering
	\includegraphics[width=\reprintcolumnwidth]{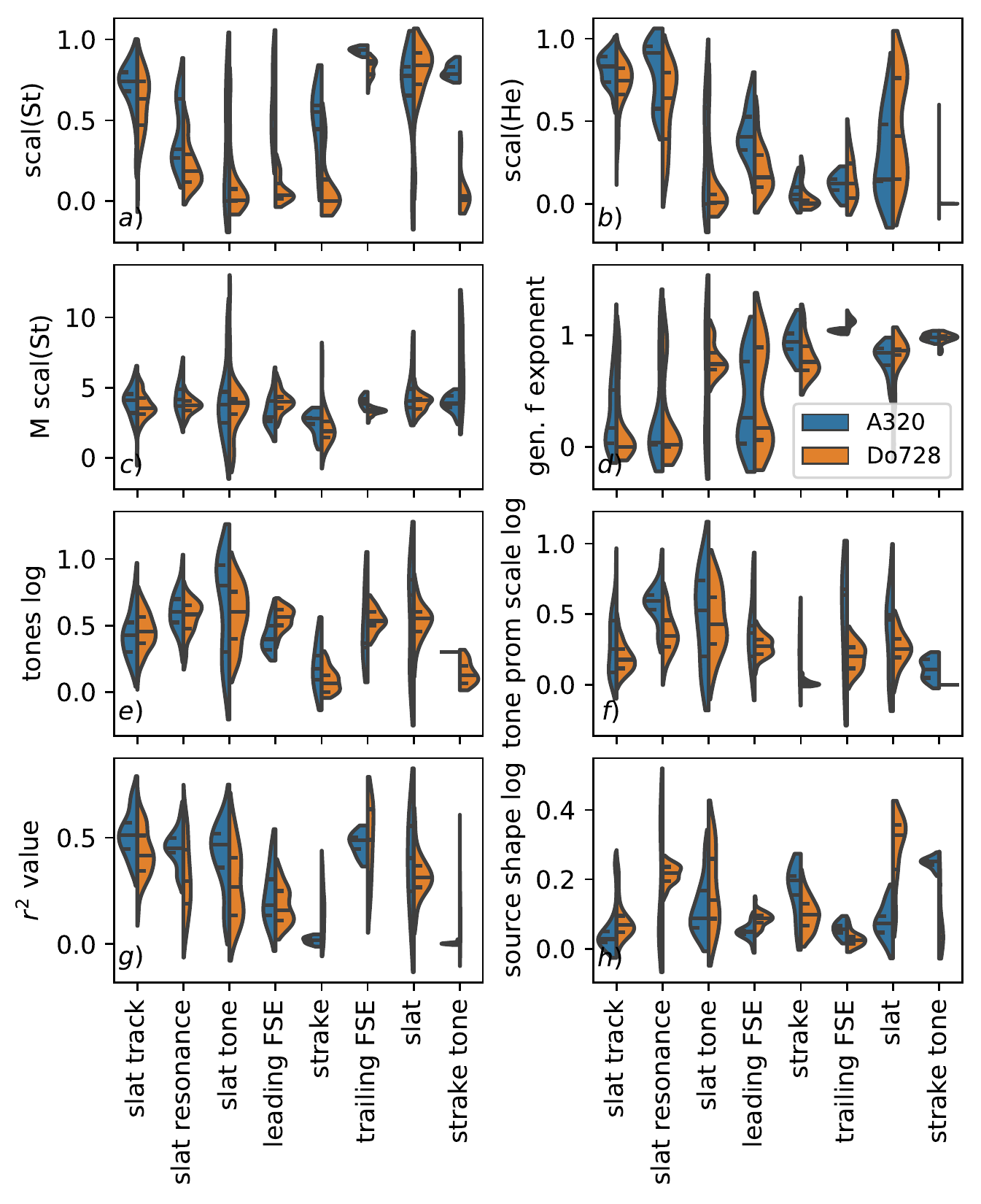}
	\caption{(Color online) Violin plot comparison of the A320 (blue) Do728 (orange) for exemplary source types and exemplary features from table~\ref{tab:features_table}. $a)$ shows the broadband self-similarity over the modified Strouhal number, see Section~\ref{sec:similarity}. $b)$ shows the broadband self-similarity over the Helmholtz number. $c)$ shows the Mach power scaling over the modified Strouhal number, see Section~\ref{sec:Mach_number_scaling}. $d)$ shows the generalized frequency modification exponent, see Section~\ref{sec:mod_exponent}. $e)$ shows the logarithmic number of tones, and $f)$ shows the gamma distribution's logarithmic scale of the tone prominence, see Section~\ref{sec:tonality}. $g)$ shows the $r^2$-value of the linear regression of the spectrum, see Section~\ref{sec:spectrum_shape}, and $h)$ shows the logarithmic spatial source shape, see Section~\ref{sec:srs_distribution}.}
	\label{fig:Figure9}
\end{figure}

Figure~\ref{fig:Figure9} presents the distributions of exemplary features for exemplary source types and compares them between both datasets. The horizontal lines within the distributions display the 0.25, 0.50, and 0.75 percentiles. The exemplary source types were chosen because they were manually identified in Section~\ref{sec:classification} with great confidence based on their spectral features in Figure~\ref{fig:Figure7} and they showed interesting acoustical properties. The features were chosen to cover the variety of aeroacoustic properties introduced in Section~\ref{sec:features}.\deleted[R1C10]{, i.e. self-similarity, Mach power scaling, generalized frequency modification exponent, tonality, spectrum shape, and spatial source shape. We present an additional analysis for features that are significantly correlated with the angle of attack and the Mach averaged Reynolds number. The number of occurrences $n_O$ of source types itself also depends on the angle of attack and Reynolds number. For the slat tracks the occurrences correlate with the angle of attack $\rho_{n_O,\alpha}=0.68$, $p=0.045$ (p-values are rounded to three-digit precision), but not with the Reynolds number $\rho_{n_O,\text{Re}}=-0.44$, $p=0.234$. The occurrences of slat tone sources does not correlate significantly with the angle of attack $\rho_{n_O,\alpha}=0.38$, $p=0.313$, but not with the Reynolds number $\rho_{n_O,\text{Re}}=-0.81$, $p=0.008$. There is a strong correlation for strake sources with the angle of attack $\rho_{n_O,\alpha}=0.82$, $p=0.006$, and with minor significance with the Reynolds number $\rho_{n_O,\text{Re}}=-0.60$, $p=0.085$. There is a strong correlation for slat sources with the angle of attack $\rho_{n_O,\alpha}=-0.93$, $p=0.000$, but no significant correlation with the Reynolds number $\rho_{n_O,\text{Re}}=-0.30$, $p=0.427$. The occurrences of slat resonance, LFSE, and TFSE sources do not significantly correlate with the angle of attack or the Reynolds number.}\\

Figure~\ref{fig:Figure9} shows the self-similarity over $a)$ the modified Strouhal number, and $b)$ the Helmholtz number, see Section~\ref{sec:similarity}. \replaced[R1C01,R2C04h]{Based on this definition, a strong Strouhal scaling behavior (scal(St)$\approx$1) is observed for the slat tracks, the TFSE, the slats, and the A320 strake tones. A weak Strouhal scaling (scal(St)$\approx$0.5) is observed for the Do728 strake. No Strouhal scaling behavior (scal(St)$\approx$0) is observed for the Do728 slat resonance, slat tones, LFSE, strake, and the strake tones. Strong Helmholtz scaling is observed for the slat tracks and the slat resonance. A weak Helmholtz scaling is observed for the LFSE. No clear trend can be observed for the Do728 slat tones. There is a correlation of the Reynolds number with the Strouhal scaling (all sources) $\rho_{\text{scal(St)},\text{Re}}=0.16$, $p=0.000$ and the Helmholtz scaling (all sources) $\rho_{\text{scal(He)},\text{Re}}=0.08$, $p=0.003$), but none with the angle of attack.} 
{Based on the exemplary source spectra in Figure~\ref{fig:Figure7} and the aeroacoustic literature we expect the slat track, slat resonance, and leading flap side edge to strongly scale over the Helmholtz number and the slat tone, strake, trailing flap side edge, slat and strake tone to strongly scale over the (modified) Strouhal number. For the slat tracks the feature does not achieve satisfying results, since it wrongly predicts a strong self-similarity over the Strouhal number for both datasets. The reason for this is the strong decay in SPL over frequency, which dominates the correlation stronger than the local spectral features such as tonal peaks or humps. For the Do728 strake tones, a low Strouhal self-similarity is wrongly predicted. This is due to the inclusion of the $p$-value in eq.~\ref{eq:scalability}, which becomes large for small-band sources. For the A320 slat tone and LFSE the feature is not robust and results in a large range of values. The expected self-similarity over the Helmholtz number is well captures in Figure~\ref{fig:Figure9} $b)$. This is also true for Strouhal number scaling slat sources with additional Helmholtz scaling high-frequency content, see Figure~\ref{fig:Figure8} $a)$ at $100\le\text{St}\le200$.}\\

Figure~\ref{fig:Figure9} $c)$ shows the Mach power scaling over the modified Strouhal number. Similar power scalings around $M$ scal(St)$\approx3.73$ (averaged over the displayed source types) are observed. \replaced[R1C01,R2C04h]{No clear trend is observed for the slat tones and strake tones. When scaled over the Helmholtz number, the average power scaling is around $M$ scal(He)$\approx5.17$. For comparison, the average OASPL scaling results in $M$ scal(OASPL)$\approx4.73$. There are no significant correlations with the Reynolds number or angle of attack.}{This is expected as aeroacoustic noise is known to scale within a small range (e.g., $M^4$ for monopoles, $M^6$ for dipoles, and $M^8$ for quadrupoles), depending on the source mechanism. However, the variance within the source types often exceed the variance between the source types. Also, as seen for the trailing flap side edge the Mach scaling can differ for the same source type for different datasets. At this point it is not clear if this is caused by the assumptions of the method (e.g., ignoring the source directivity) or by the different model geometries.}\\

Figure~\ref{fig:Figure9} $d)$ shows the generalized frequency normalization exponent $m$, see Section~\ref{sec:mod_exponent}. Most of the slat tracks and slat resonances show an exponent around $m\approx0$. This indicates a Helmholtz scaling, and corresponds to the presented self-similarities. For both datasets there are two groups of LFSE, one that features $m\approx0$, and one that features $m\approx1$. \deleted[R1C10]{For the A320 there is a strong correlation between the exponent and the angle of attack $\rho_{m,\alpha}=-0.69$, $p=0.003$, but none for the Do728 $\rho_{m,\alpha}=0.10$, $p=0.546$. There is no significant correlation with the Reynolds number.} For the Do728 strakes a mean exponent $m\approx0.79$ is observed, for the A320 strakes $m\approx0.95$. \deleted[R1C10]{There is a strong correlation with the angle of attack with $\rho_{m,\alpha}=-0.47$, $p=0.005$ for the Do728 and $\rho_{m,\alpha}=-0.95$, $p=0.050$ for the A320. There is no significant correlation with the Reynolds number.} For the Do728 TFSE $m\approx1.13$ is observed, for the A320 $m\approx1.06$. \deleted[R1C10]{There are no significant correlations with the angle of attack. For the A320 there is a correlation with the Reynolds number $\rho_{m,\text{Re}}=0.61$, $p=0.012$, but none for the Do728 $\rho_{m,\text{Re}}=0.10$, $p=0.555$.} For the Do728 slat $m\approx0.78$ is observed, for the A320 $m\approx0.72$. \deleted[R1C10]{There is a correlation between the Do728 exponent and the angle of attack $\rho_{m,\alpha}=-0.37$, $p=0.007$, but not with the Reynolds number $\rho_{m,\text{Re}}=0.17$, $p=0.233$. There are no correlations for the A320 slat sources.}\added{These results show that source mechanisms can increase and decrease the Strouhal numbers Mach dependence.}\\

Figure~\ref{fig:Figure9} $e)$ shows the Mach averaged occurrence of tones (logarithmic), see Section~\ref{sec:tonality}. Some source types feature few tones, i.e. the strake and strake tone, some source types feature many tones, i.e. the slat tones. Figure~\ref{fig:Figure9} $f)$ shows the corresponding scale parameter of the tone prominence distribution (logarithmic). A large scale parameter indicates that the distribution is spread out, including tones with small and large prominence. A small scale parameter indicates that all tones have a similar prominence. Both datasets show similar distributions for the features. \deleted[R1C10]{For the Do728 slat tones there is a correlation with the Reynolds number $\rho_{\text{tones},\text{Re}}=-0.37$, $p=0.008$ and for the A320 slat tones $\rho_{\text{tones},\text{Re}}=-0.38$, $p=0.046$. The Do728 slat tones' tone prominence scale parameter $\theta$ is correlated with the Reynolds number $\rho_{\theta,\text{Re}}=-0.37$, $p=0.008$, and for the A320 $\rho_{\theta,\text{Re}}=-0.62$, $p=0.000$.} \added{Generally, the features correspond to our expectations. However, the gamma distribution approximation fails for sources which contain only one prominent tone, such as the Do728 strake tone sources.}\\

Figure~\ref{fig:Figure9} $g)$ shows the spectrum shape $r^2$ based on a linear regression, see Section~\ref{sec:spectrum_shape}. For the Do728 the distributions of the strake, the TFSE, and the strake tone are spread out compared to the A320 distribution. Overall, they show similar trends. \deleted[R1C10]{There are no significant correlations with the Reynolds number or angle of attack.}\added{The feature mainly highlights the fact, that many aeroacoustic sources are broad-band sources with a linear decay in SPL over logarithmic frequency and correctly identifies small-band sources such as the strake and strake tone.}\\

Figure~\ref{fig:Figure9} $h)$ shows the source shape (logarithmic). A value close to zero indicates a point source, an increasing value indicates a line source. \deleted[R1C10]{There are large deviations between both datasets for the slat resonance, the slat, and the strake tone. There are no significant correlations with the Reynolds number or angle of attack.} \added{The results directly depend on the output of SIND, which correctly identified the slat track, LFSE, strake, and TFSE as point-like sources. The Do728 slat was correctly identified as a line-like source. However, the A320 slats were wrongly identified as point-like sources. While SIND yielded overall comprehensible results, on some occasions sources were wrongly spatially separated or combined. E.g., the slat tones sometimes appear on slat track positions and thus, are identified as point-like sources, while they should be line-like sources~\citep{Dobrzynski2001}.}
\\

\begin{figure}
	\centering
	\includegraphics[width=\reprintcolumnwidth]{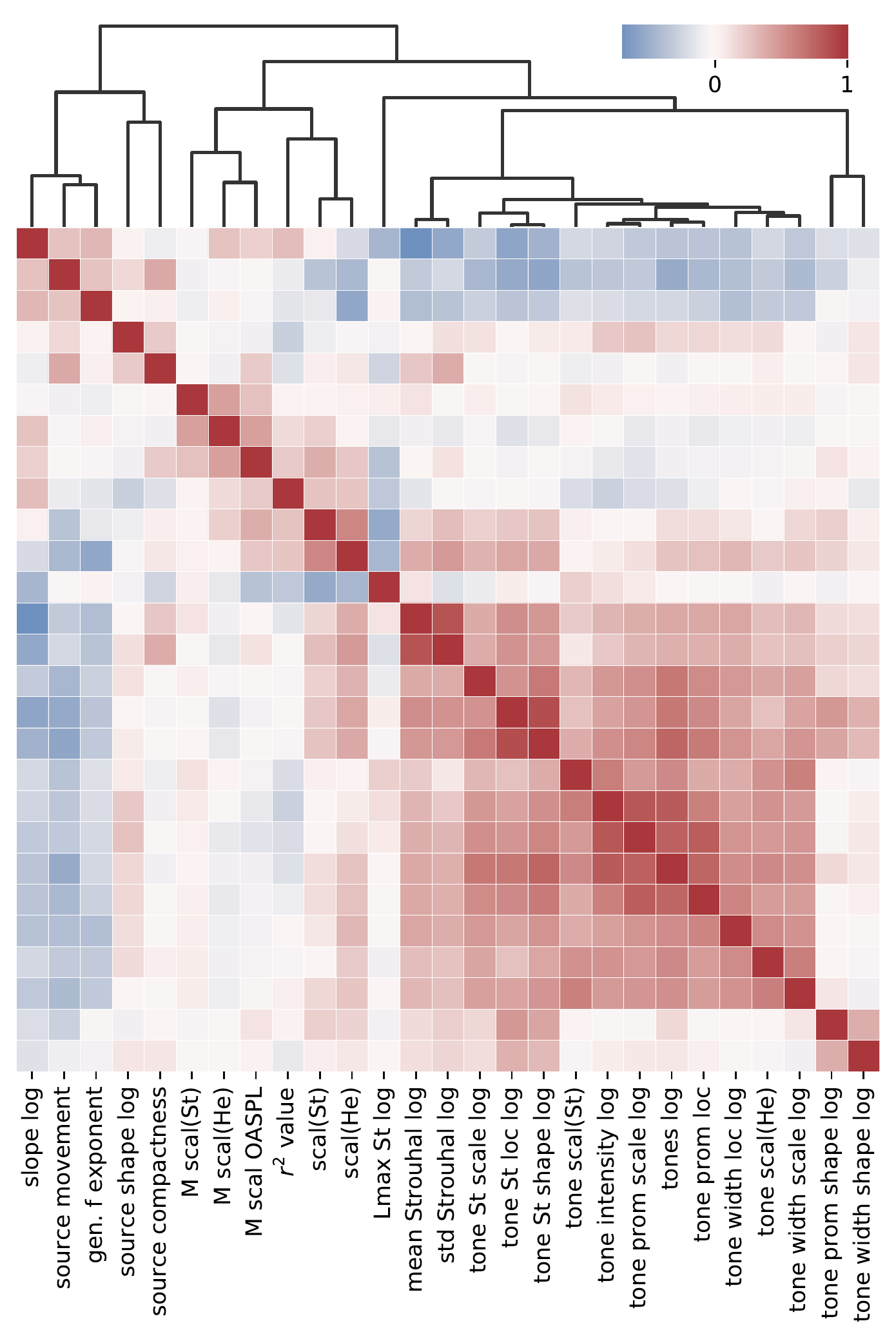}
	\caption{(Color online) Heatmap of all Do728 feature Pearson correlation coefficients and their hierarchy based on hierarchical clustering with a correlation distance metric (top).}
	\label{fig:Figure10}
\end{figure}

Figure~\ref{fig:Figure10} shows a \deleted{heatmap of }Pearson correlation coefficient \replaced{matrix}{heatmap} for the Do728 feature space. On the top, a feature hierarchy is displayed based on hierarchical clustering with a correlation distance metric based on the Unweighted Pair Group Method with Arithmetic mean (UPGMA) algorithm. This hierarchy shows which features are similar to each other based on the displayed correlations to all other features. The heatmap and hierarchy show that many of the introduced features correlate strongly, especially features that were introduced together to cover an aeroacoustic property such as tonality, or self-similarity. The feature hierarchy shows that these features originate from the same branch. From left to right, the first major branch includes the linear regression's slope (log), the source movement, the generalized frequency exponent, and the source compactness. The first three features correlate negatively with the features that correspond to the tonality and form a sub-branch. The second main branch on the right contains all other features. Its left sub-branch contains the Mach scaling over Strouhal number, and Helmholtz number, as well as the linear regression's $r^2$-value, and the self-similarity (scal) over Strouhal number and Helmholtz number. The other branch contains all features that describe the frequency content of the spectrum (St$_{L\text{max}}$, mean and std Strouhal numbers), and its strongly correlated tonality features.\\

\begin{figure}
	\centering
	\includegraphics[width=\reprintcolumnwidth]{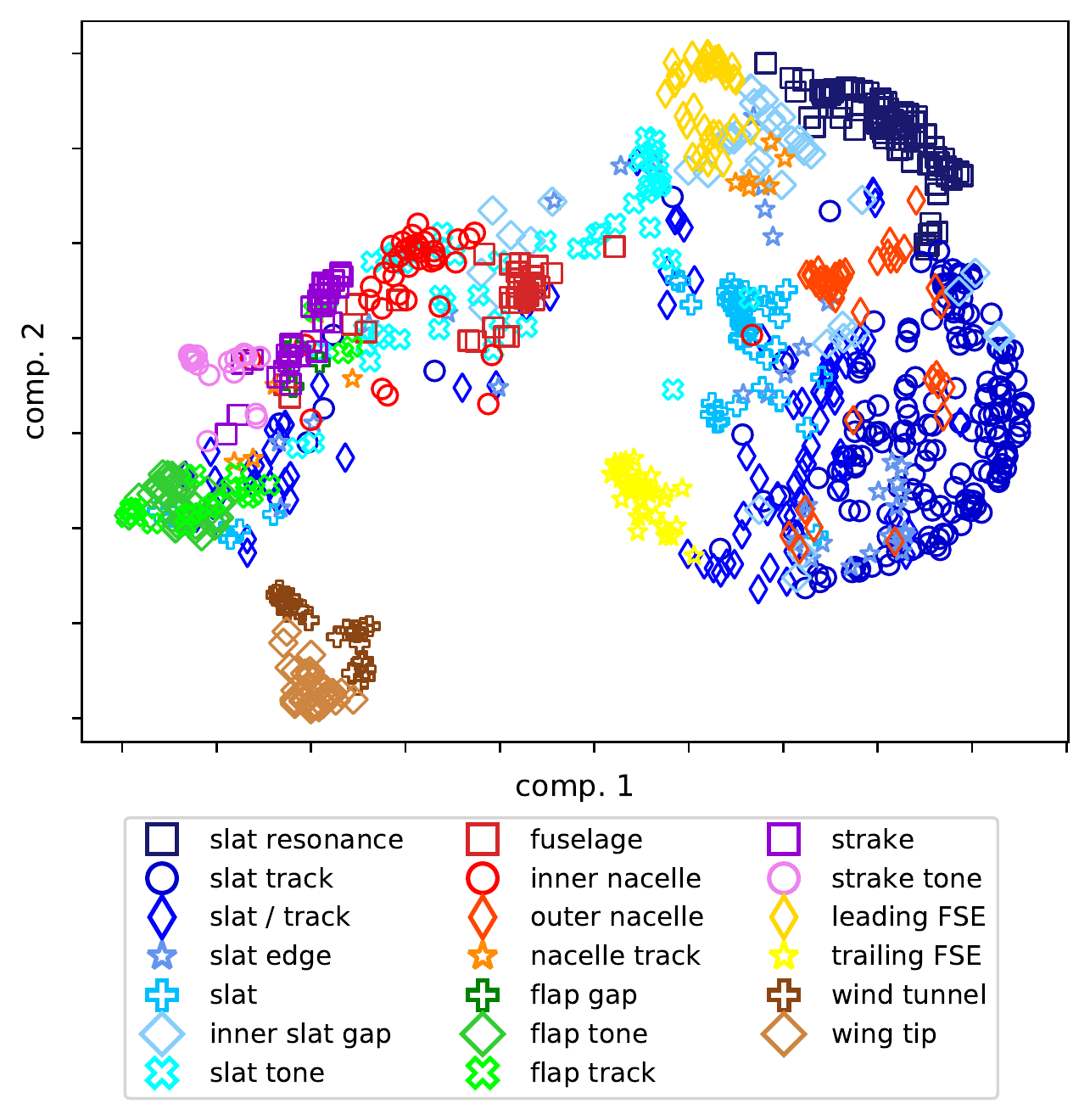}
	\caption{(Color online) Do728. UMAP~\citep{Mcinnes2020} for $n_\text{components}=2$, $n_\text{neighbors}=90$ ($\approx\SI{10}{\percent}$ of the dataset), and a correlation distance metric for the presented feature space.}
	\label{fig:Figure11}
\end{figure}

Figure~\ref{fig:Figure11} shows a Uniform Manifold Approximation and Projection for Dimension Reduction~\citep{Mcinnes2020} (UMAP) of the Do728 sources. The colors and symbols both represent the manual labels. UMAP estimates a manifold that locally (and to some extend globally) preserves the data structure in a low dimensional space, based on which the data can be displayed in low-dimensional space. UMAP requires a distance metric, such as a spatial metric (e.g., Mahalanobis, Minkowski), or a similarity metric (e.g., cosine, correlation). All mentioned metrics yield similar results for the presented feature space. Figure~\ref{fig:Figure11} shows a 2D space, in which we can observe how similar source types are in the introduced feature space\added[R1C11]{ and how well the introduced feature space separates the proposed source types}. \replaced[R1C11]{In the middle of the image, there are the trailing flap side edge sources (yellow star). There are mostly separated from other groups and slightly touch the ambiguous slat / track category (blue diamond) on the right which includes both slat and slat track sources. This group lies in the middle of the slat tracks and some slat edges (bottom right, blue circle), and the slats (top, blue plus). On the right of the slat sources. The closest category is the outer nacelle (top, red diamonds). On the top right, there is the slat resonance (blue square), which connects to the slat track group on the bottom and the inner slat gap (blue rotated square), and the nacelle track (red star) on the left. These sources connect to the leading flap side edge (left, yellow diamond), and then the slat tones (left, blue x), which also connects to the slat sources. The next sources are the fuselage (left, red square), the inner nacelle (left, red circle), strake (bottom left, violet square), strake tone (left, violet circle), flap tone (bottom, green rotated square), and flap track (green x). Slightly separated is the wind tunnel (bottom, brown plus), and the wing tip (bottom, brown rotated square). Additional outliers are scattered throughout the projection, especially slat, slat track, and slat / track sources.}{The slat track and slat edge sources are similar, and gradually transition to slat / track sources, and then to slat sources. These groups are not well separated which corresponds to our own assessment, since the spectra often smoothly transition from one shape to another and show great variance. The slat tone sources are not captured well in the feature space, since they are manually identified with very high confidence, but do not form a well separated group. Other source types such as the wing tip and the trailing flap side edge are well captured in the feature space and show little variance. Overall, the Figure shows that sources of the same manually introduced source type are close to each other in the introduced feature space. Some outliers, which are not close to their core group, are also visible. There are three possible explanations for this behavior. First, the manual labeling is wrong or ambiguous. Second, the automated feature calculation resulted in wrong values due to an insufficient definition or because the formula was not robust enough towards the degenerated spectra. Third, the spectra contained partially wrong information due to the limitations of the SIND process~\citep{Goudarzi2020}. The ability of UMAP to form separable source type groups highlights two results. First, the introduced feature space captures sufficient aeroacoustic information for the presented sources, under the condition that the labeling is correct. Thus, groups of multiple sources are formed in the UMAP based on the feature space. Second, the labeling is sufficiently good under the condition that the feature space correctly captures the sources' aeroacoustic behavior. Thus, the groups mainly contain unique source types. Since the source types were identified manually based on the source spectra and not on the introduced feature values, see Section~\ref{sec:classification}, this indicates that the label choices are reasonable.}\\

\subsection{Clustering results}

\begin{figure}
	\centering
	\includegraphics[width=\reprintcolumnwidth]{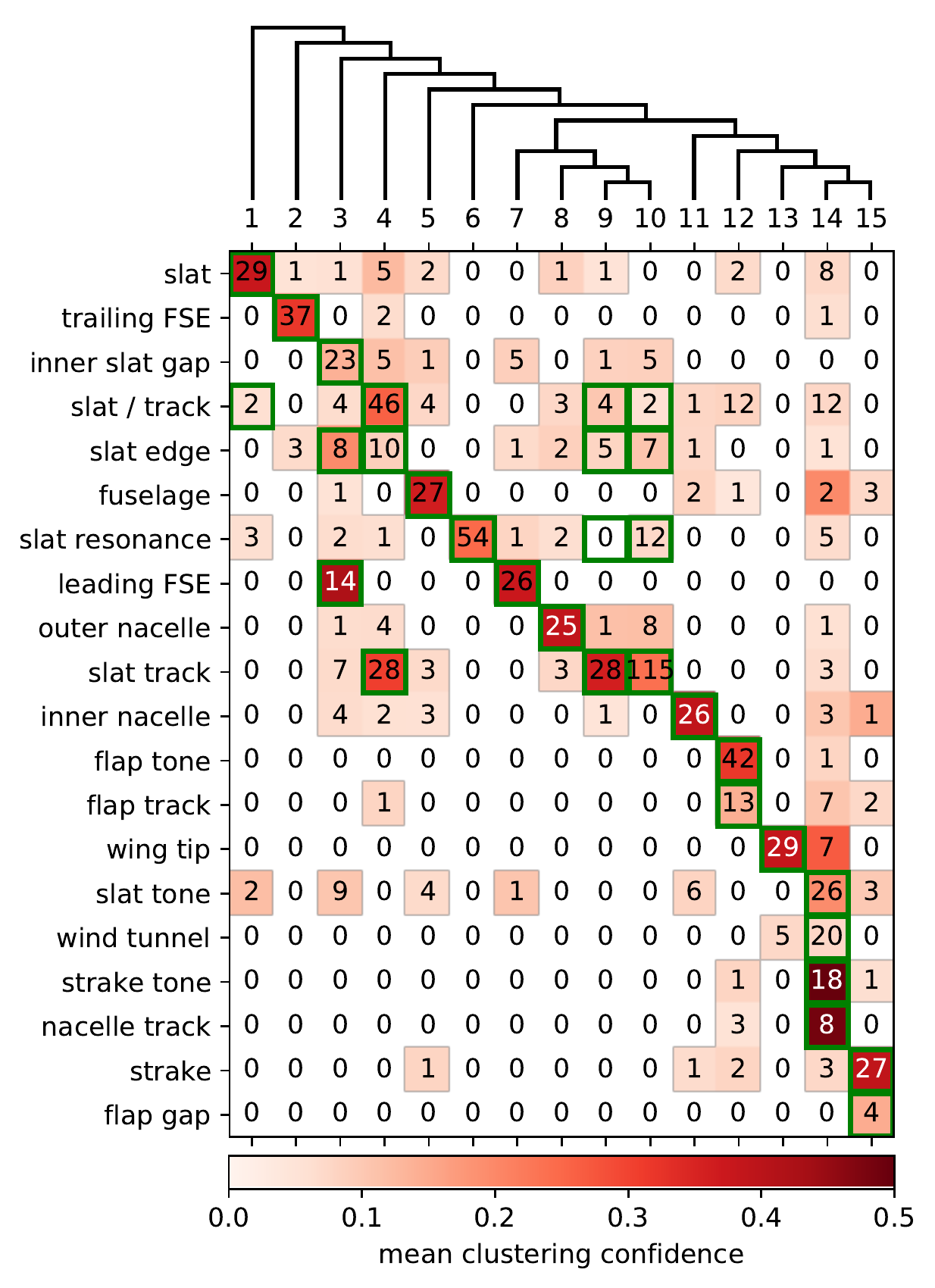}
	\caption{(Color online) Do728, confusion matrix of the occurrences of our manually identified source types in CRAFT's determined clusters. The color intensity displays the mean clustering confidence. Cluster choices we consider correct are marked with a green box. The tree above the clusters displays their hierarchy.}
	\label{fig:Figure12}
\end{figure}

\begin{figure}
	\centering
	\includegraphics[width=\reprintcolumnwidth]{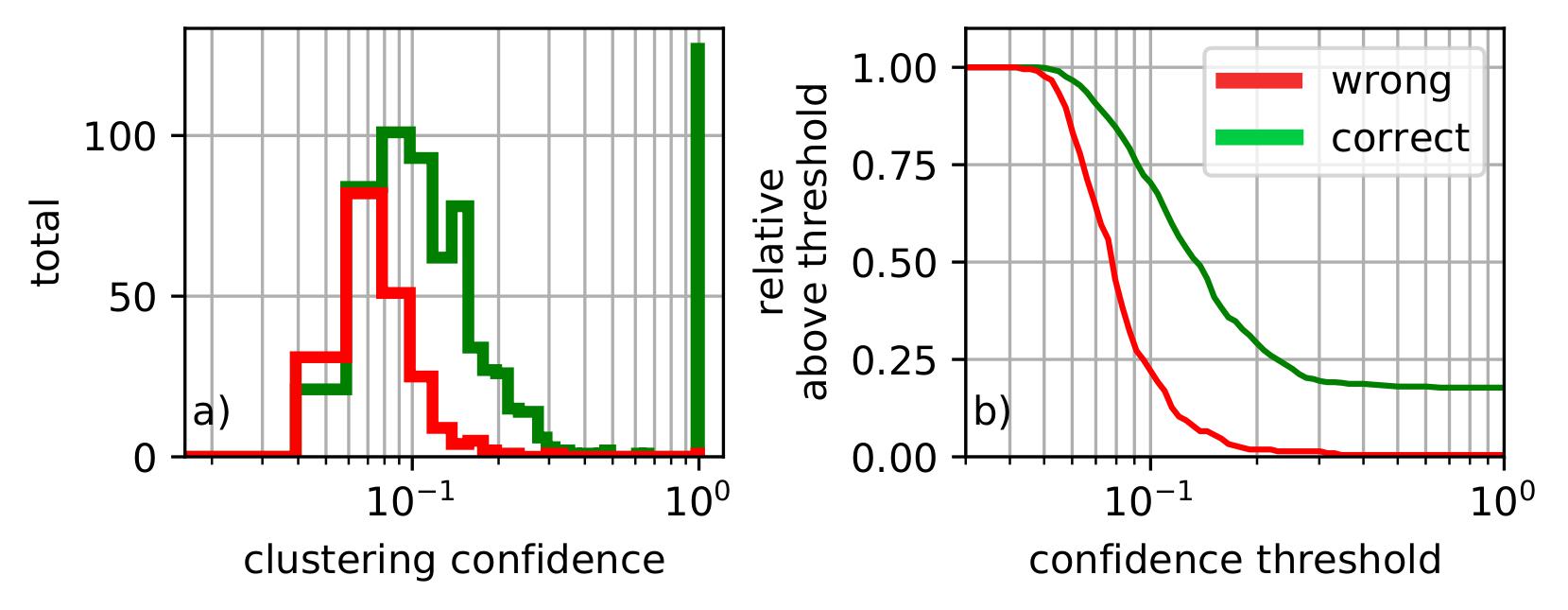}
	\caption{(Color online) Do728, $a)$ shows a histogram of the total wrong and correct cluster choices over their clustering confidence. $b)$ shows the relative number of cluster choices above the confidence threshold $t_C$ on the $x$-axis.}
	\label{fig:Figure13}
\end{figure}

The results presented are based on the Do728 and A320 datasets, obtained from sparse beamforming maps with SIND and CRAFT. For the Do720, a minimum sample size of seven was used for HDBSCAN. CRAFT determined fifteen source clusters. Figure~\ref{fig:Figure12} shows a confusion matrix of the manually determined labels, see Section~\ref{sec:classification}, and the clustering results. The matrix shows how often a source from a manually determined category was clustered into the corresponding clusters. Given the manual source type identification is correct, a perfect clustering would result in a cluster group for every manual label and all corresponding sources would be clustered within their corresponding group. Thus, a perfect clustering would achieve a square confusion matrix with all results on the diagonal. The underlying color in the confusion matrix depicts the mean clustering confidence. Since most clusters correlate to the manual labels (they are mostly located on the diagonal axis), our definition of correct clustering results (which are marked with a green box in the confusion matrix) will be based on its comparison to the manual labels.\\ 

To identify the clusters which correspond to the manual labels we will take the occurrences of source types per cluster, their estimated confidence, their similarity to other source types, and the cluster hierarchy into account. Thus, cluster number one is assigned to the slat sources. We consider slat sources that are assigned to cluster one as correct, slat sources assigned to other clusters are considered wrong. A slat resonance which was categorized as cluster number nine or ten (which consists mainly of slat tracks) instead of cluster number six (which consists solely of slat resonances) is considered correct since it is a sub-type of a slat track source. All sub-categories that are clustered with their super-categories are considered correct, but not vice versa. Thus, slat tracks that are clustered in group six (slat resonances, a sub-category of slat tracks) are considered wrong. Slats and slat tracks that fall in cluster four (slat / slat track) and vice versa are considered correct, which is a super-category of these ambiguous sources. Cluster numbers three, four, fourteen, and fifteen comprise multiple source types. As long as the corresponding source types were assigned to the cluster that contained most of the sources, they are considered correct. The slat tracks occupy the two clusters number nine and ten with high clustering confidence, which we consider as equally correct, as the clusters are both branches of a super-cluster. The leading flap side edge occupies both cluster numbers three and seven with high confidence, thus we consider both clusters as correct. All other clustering results are considered as wrong.\\

Note that the following clustering assessment is based on both our manual source labeling and our manual definition of correct confusion matrix entries. In total, 213 out of 928 Do728 source predictions ($\SI{22.95}{\percent}$) are considered wrong and 715 ($\SI{77.04}{\percent}$) are considered correct. Figure~\ref{fig:Figure13} $a)$ shows the number of clustering choices at the given clustering confidence. Both, the correct and wrong clustering choices decrease with increasing confidence. Figure~\ref{fig:Figure13} $b)$ shows the relative number of wrong and correct clustering choices that lie above the confidence threshold $t_C$. We observe that the wrong clustering results decrease much more rapidly than the correct clustering results. As an example, if the clustering results with confidence below $t_C=0.1$ are discarded, only a prediction for $\SI{59.26}{\percent}$ of the sources is retained, but the clustering accuracy increases to $\SI{91.45}{\percent}$.\\

\begin{figure}
	\centering
	\includegraphics[width=\reprintcolumnwidth]{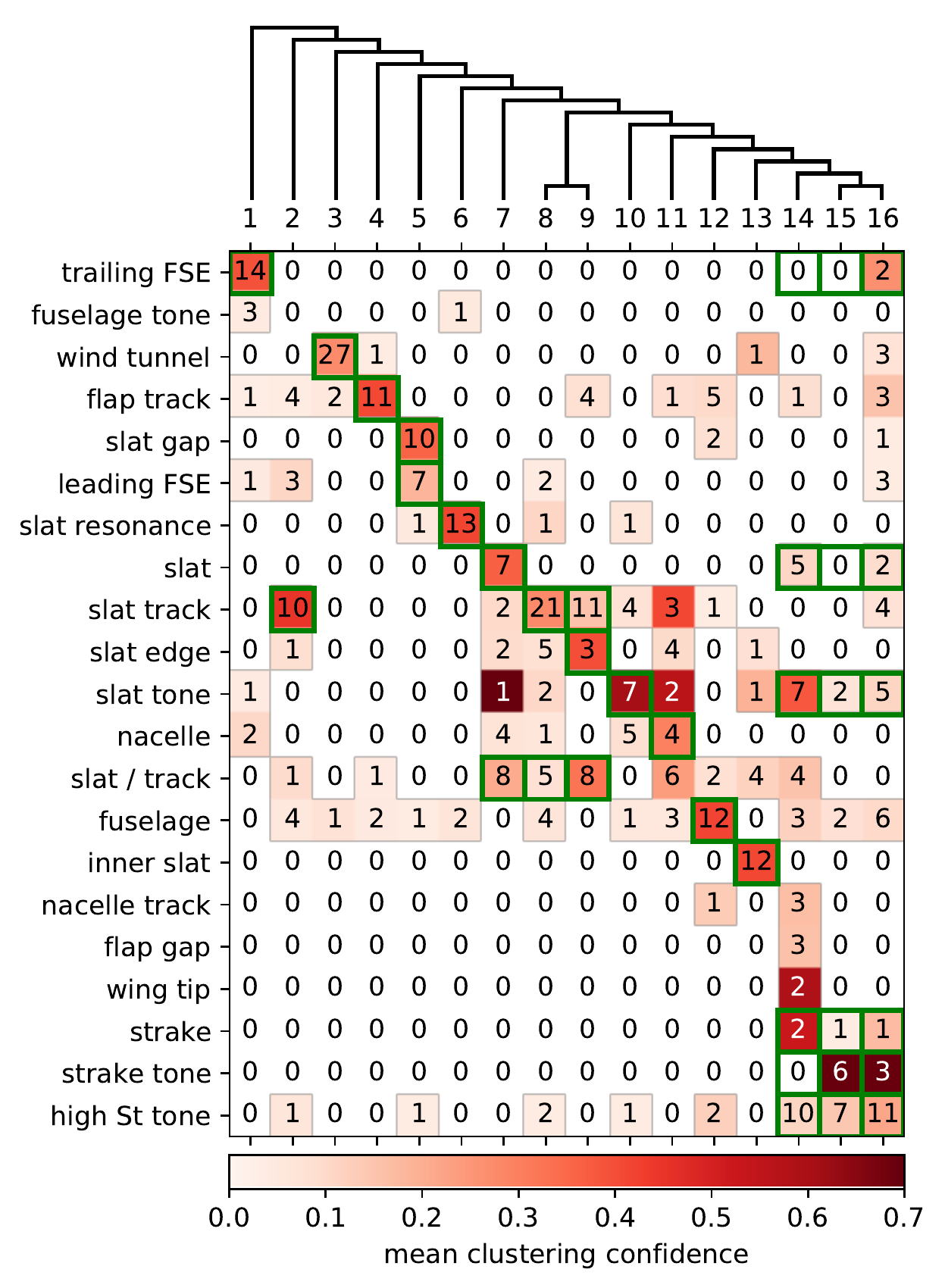}
	\caption{(Color online) A320, confusion matrix of the occurrences of our manually identified source types in CRAFT's determined clusters. The color intensity displays the mean clustering confidence. Cluster choices we consider correct are marked with a green box. The tree above the clusters displays their hierarchy.}
	\label{fig:Figure14}
\end{figure}

Figure~\ref{fig:Figure14} shows the confusion matrix of our manually identified source types and the clustering results for the A320. The correct clusters are determined as stated above for the Do728. In particular, the super-cluster 14, 15, and 16 represent Strouhal number scaling tones. Thus, we decided to exclude the nacelle track, flap gap, and wing tip from the correct results. The super-cluster 7,8, and 9 represent slat / tracks, with sub-cluster 7 including the slats and sub-clusters 8, and 9 including the slat tracks and the slat edge. In total, 154 out of 408 source predictions ($\SI{37.75}{\percent}$) are wrong and 254 ($\SI{62.25}{\percent}$) are correct. As shown for the Do728 \replaced{classification}{results}, if the clustering choices with confidence below $t_C=0.1$ are discarded, only a prediction for $\SI{51.47}{\percent}$ of the sources is retained, but the clustering accuracy increases to $\SI{74.29}{\percent}$.\\

%%%%%%%%%%%%%%%%% DISCUSSION %%%%%%%%%%%%%%%%%%%%%%%%%%%%%%%%%%%
\section{Discussion}
\deleted[R1C01,R2C04h]{We presented the EDSS CRAFT and evaluated its results on two datasets, based on CLEAN-SC beamforming maps of two similar scaled airframe models in closed wind tunnel sections, an A320 and a Do728. }CRAFT is built on the assumption that an aeroacoustic source is driven by a mechanism that reveals its nature over the variation of the Mach number. Thus, we used spectra at different Mach numbers to determine the source's aeroacoustic properties. The limitation of this general assumption is that we neglect any changes of the source mechanism over the Mach number. We showed in Figure~\ref{fig:Figure3} $b)$ at high Helmholtz numbers that sources can exhibit a substantial Mach dependency, which is not captured by the proposed method. This means that Mach number dependent source mechanisms cannot be detected by the proposed method. However, the advantage of this assumption is that the properties are defined independently of the employed Mach numbers, which allows the clustering and comparison of data from different measurement configurations.\\

% features
Based on this assumption, we proposed a variety of features in Section~\ref{sec:features} that are supposed to formalize aeroacoustic properties and that are robust towards degraded spectra. This is necessary since we used CLEAN-SC beamforming maps in combination with the SIND method~\citep{Goudarzi2021} to automatically extract the source spectra, their positions, and their spatial distribution. The resulting spectra are often not very reliable at low frequencies or contain missing values. This is especially problematic since their dominant SPL, which is preferred in many features, occurs at these low frequencies.\deleted[R1C01,R2C04h]{ The presented properties were selected to capture common aeroacoustic concepts and to explain the different source types in the presented datasets. We proposed formulas for the features with the main goal of robustness during their automated calculation.} We neither claim that the list of the properties is complete and covers all acoustic phenomena nor do we claim that the feature calculation is robust in all data scenarios. We hope to spark a discussion in the acoustic community on which properties are important for which source types, how these can be broken down to numeric feature values, and how they can be calculated robustly and efficiently.\\

% self-similarity
In sub-section~\ref{sec:similarity}, we presented a calculation method to determine how self-similar the source spectra are over a given frequency type, i.e. the Helmholtz and Strouhal number. \deleted[R1C01,R2C04h]{The self-similarity is based on the Pearson correlation coefficient between spectra at different Mach numbers. The calculation is irrespective of the number of different Mach numbers in the dataset (however, the calculation accuracy increases with more data) and results in a soft value $-1\le\text{scal}\le1$. To account for degenerated spectra (missing values in the spectra, or multiple source mechanisms within a spectrum) we included the standard deviation of the correlations and the p-value in eq.~\ref{eq:scalability}. Exemplary spectra for source types were given in Figure~\ref{fig:Figure7}, Figure~\ref{fig:Figure9} $a)$ and $b)$ showed a comparison of the resulting self-similarities per source type of the A320 and Do728.} For \replaced{the slat tracks}{several source types}, the \replaced{method}{feature} does not achieve satisfying results. E.g., the self-similarity wrongly predicts a strong self-similarity over the Strouhal number for both datasets. The reason for this is the strong decay in SPL over frequency, which dominates the correlation stronger than the local spectral features such as tonal peaks or humps. In the corresponding conference paper~\citep{Goudarzi2020} we used the variance over frequency of the \replaced{distance between the}{variance of} spectra at different Mach numbers to calculate the self-similarity. This also allows a weighting of the \replaced{distances}{variances} with the Mach averaged SPL. However, this \replaced{method}{definition} wrongly predicts a low self-similarity for sources like the slat tones, since the tones are aligned but vary greatly in SPL.\replaced{For the majority of slat resonances, the current definition correctly predicts a strong self-similarity over the Helmholtz number and a weak self-similarity over the Strouhal number. For some sources, there is also an elevated self-similarity over the Strouhal number, which can be explained by the high-frequency content that indeed scales over the Strouhal number, see Figure~\ref{fig:Figure8} $n)$ at $25\le\text{He}\le50$. The \replaced[R1C34]{LFSE}{leading flap side edge} and slat tones are additional examples where the presented method fails to correctly predict the self-similarity. For the \replaced[R1C34]{TFSE}{trailing flap side edge} and the slat, the results correctly capture the strong Strouhal scaling behavior. The additional high-frequency content in the slat sources that scale over Helmholtz number, see Figure~\ref{fig:Figure8} $a)$ at $100\le\text{St}\le200$, is also well captured in the corresponding self-similarity. Including the p-value in the self-similarity was a necessary step to determine satisfying results for the generalized frequency exponent. However, this results in unsatisfying self-similarities for small-band sources like the strake tone, since the p-values of the correlations become very large, and thus, the self-similarity is small.} We encourage the community to propose a mathematical definition that fixes these issues.\\

% mod freq exponent
In sub-section~\ref{sec:mod_exponent} we showed that the normal Strouhal number definition is not sufficient when working with source spectra at different Reynolds numbers. \replaced[R1C01,R2C04h]{Since $\text{Re}\propto M$ at constant temperature and ambient pressure, the Reynolds number typically increases with increasing Mach number within a measurement configuration. This can result in a shift of the peak Strouhal number over the Mach number (the Helmholtz number does not depend on the Mach number and thus, does not shift). To account for this}{Thus}, we defined a normalized frequency that is a generalization of the Strouhal and Helmholtz number by introducing the modification exponent $m$. We found \replaced{for noise that is related to the slat or slat tones a modification exponent of $m\approx 0.76\pm 0.08$ and for the trailing flap side edge $m\approx 1.10\pm 0.02$ (averaged over both datasets). Thus,}{that} some source mechanisms increase their frequency dependency on the Mach number, while others decrease their Mach dependency. \deleted[R1C01,R2C04h]{Even though this drift is caused by the increasing Reynolds number, no correlation between the exponent and the Mach averaged Reynolds number itself was found. This is interesting since the high Reynolds numbers are produced by high pressures and low temperatures, see Table~\ref{tab:A320}. At these conditions, the increase in Reynolds number with increasing Mach number is much stronger than at ambient conditions, which should result in a stronger drift of the frequency and thus, a correlation. }To our knowledge, this is the first time this phenomenon is described, but it is too complex to be fully covered within the scope of this paper. More research is necessary to understand the underlying acoustic mechanisms and the implications on wind tunnel measurements.\\

% power exponent
In sub-section~\ref{sec:Mach_number_scaling} we introduced a method to determine the power exponent \added{for a seperate scaling over the Strouhal and Helmholtz number} using a weighted standard deviation between the spectra, see eq.~\ref{eq:Mach_scaling}. \replaced[R1C01,R2C04h]{The method aims to minimize the SPL-weighted distance of the scaled spectra. In Figure~\ref{fig:Figure7} we presented the scaling results for exemplary source spectra. We found that a global hyperparameter $6\ge\gamma\ge8$ works well for the regularization in eq.~\ref{eq:Mach_scaling} (penalty for low SPL). In Figure~\ref{fig:Figure9} $c)$ we showed the resulting power exponent distributions, which are very similar. Aeroacoustic noise is known to scale within a small range (e.g., $M^4$ for monopoles, $M^6$ for dipoles, and $M^8$ for quadrupoles), depending on the source mechanism. In Figure~\ref{fig:Figure10} we showed that the power scaling over Strouhal number, Helmholtz number, and OASPL are correlated and belong to the same branch of the feature hierarchy, next to the self-similarity. This suggests that the scaling behavior is indeed a unique aeroacoustic property.}
{However, since we assumed only small changes in the Mach number, and based on the small amount of observed Mach numbers, the correct determination of the power exponent is difficult which causes large variances within source types, which are larger than the variances between different source types.}\\

% tonality
In sub-section~\ref{sec:tonality} we introduced a variety of features to describe the tonality of the spectra. \deleted[R1C01,R2C04h]{to overcome the challenge that the number of tones and their properties varies between spectra. }Figure~\ref{fig:Figure10} shows that the tonality features are highly correlated and belong to the same branch of the feature hierarchy. At this point, it is not clear if the tonal behavior can be captured by fewer features with less correlation. For the slat tones, the number of tones showed a significant negative correlation with the Reynolds number which suggests that this phenomenon occurs for model-sized experiments with miss-matched Reynolds numbers, but does not occur at slats at flight Reynolds numbers~\citep{Dobrzynski2001}.\\

% source movement
In sub-section~\ref{sec:srs_mov} we introduced a method to detect if sources move with increasing Mach number and in sub-section~\ref{sec:srs_distribution} a method to detect if the source is point-like or line-like\deleted{ (source shape)}. These features can only be used in combination with beamforming or other acoustic imaging methods. We used the SIND~\citep{Goudarzi2021} method to extract this information from the beamforming map\replaced{. Thus, the feature calculation relied on the outcome of the SIND method. The}{with the} main limitation \replaced{of the SIND method is that}{that the method} is tailored towards point-like sources, and that it is limited by the array resolution.\deleted[R1C01,R2C04h]{Figure~\ref{fig:Figure9} $h)$ showed that SIND correctly identified the slat track, LFSE, strake, and TFSE as point-like sources. The Do728 slat was correctly identified as a line-like source. However, the A320 slats were wrongly identified as point-like sources. Also, to be able to manually analyze these large datasets, we used SIND to find global source positions (for all Reynolds numbers, angles of attack, and Mach numbers). While this yielded overall comprehensible results, on some occasions sources were wrongly spatially separated or combined. E.g., the slat tones sometimes appear on slat track positions and thus, are identified as point-like sources, while they should be line-like sources~\hbox{\citep{Dobrzynski2001}}.} Figure~\ref{fig:Figure5} $b)$ showed the resulting source movements $\Delta l$ from eq.~\ref{eq:srs_mov}. The main observation is, that $\Delta l$ increases with the increasing source number, which is directly connected to the significance of the source due to SIND's iterative source identification. This indicates, that the feature currently estimates the position uncertainty instead of the true movement of sources (e.g. downstream moving vortex detachments). Figure~\ref{fig:Figure10} showed that the source movement, the source shape, and the source compactness are mostly correlated and that they belong to the same branch in the feature hierarchy. This suggests that they are subjected to an underlying, yet unknown, phenomenon.\\

% spectrum shape
In sub-section~\ref{sec:spectrum_shape} we introduced features to determine the general shape of the spectra, and their frequency content. Figure~\ref{fig:Figure10} showed that the\deleted{ ``waviness'' of a spectrum (the shape's} $r^2$-value is closely related to \replaced{its}{the} self-similarity and scaling behavior, \replaced{their}{the} slope is rather connected to the spatial source shape, and \replaced{their}{the} frequency content (mean and std Strouhal number) is connected to the tonality. This suggests that the current definition of the spectrum shape does contain source type-dependent variance, but does not capture a basic aeroacoustic property.\\
 
% mod freq VS st and he number
Most of the aeroacoustic properties depend on the employed normalized frequency, e.g., to calculate the power scaling exponent we need to know if the spectra are self-similar over the Helmholtz or the Strouhal number, see Figure~\ref{fig:Figure3}. This results in two options for the feature calculation. First, the correct scaling behavior is calculated and then used for the calculation of the other features. Second, the features are always calculated for spectra displayed over both the (modified) Strouhal and Helmholtz number. We decided to calculate all features that are frequency-dependent over both the Helmholtz and the Strouhal number independently. \added[R1C35]{One the one hand,} this approach proved to be robust towards spectra, that feature multiple frequency regions with different self-similarities. Additionally, the procedure is intended to support aeroacoustic experts which are already used to display spectra over both the Helmholtz and the Strouhal number. \added[R1C35]{On the other hand,} this resulted in highly correlated \replaced[R1C35]{features}{feature pairs}, see Figure~\ref{fig:Figure10}, and made a feature orthogonalization and dimensionality reduction necessary. We tried a PCA and KPCA with multiple kernels with similar results.\\

% manual classification
To evaluate the quality of the features and CRAFT, we analyzed the presented datasets manually. Due to the beamforming and \added{SIND's} spectra reconstruction process, many spectra are degenerated with missing and insufficient information. Additionally, we showed in Figure~\ref{fig:Figure11} that sources can gradually shift their spectrum shape with increasing Reynolds number and that they can feature multiple frequency regions with different mechanisms, see Figure~\ref{fig:Figure8}. Thus, we heavily relied on meta-information for the manual source type identification such as the source position. Since the proposed method was designed to overcome exactly this problem, the resulting metrics should be evaluated with caution. Also, the labeling was not performed by independent researchers.\\

% The feature distributions varied greatly for the manually identified source types. We assume the reason for this are the degenerated spectra, the fact that many spectra are a combination of different source mechanisms, and that the quantity of source type members varies greatly. We observed this in the feature projection in Figure~\ref{fig:Figure11} $b)$, where the slat tracks were scatted over the map and bridged multiple other source clusters. Nevertheless, many of our manually classified Do728 source types were separable in their feature space, independently of their location, angle of attack, Reynolds number, or SPL. This suggests that the proposed feature space included enough information to separate multiple source types and mechanisms.\\

% UMAP
To assess the combined quality of the feature space and of the manual labels we performed the dimensionality reduction and manifold estimation (UMAP) in Figure~\ref{fig:Figure11} to visualize the data in two dimensions. The \replaced{projected feature space was able to group sources of the same manually identified source types based on the introduced feature space}{projection showed that sources of the same type are mostly close in the introduced feature space}. The distance between the source groups, their distribution density, and their connections relate well to our observations in Section~\ref{sec:classification}.\deleted{ The slat tracks show the greatest variance and connect to the slat category with the ambiguous slat /track category in between.} Also, the source type distribution emphasizes the problem of manually labeling the sources based on their spectra, because source types gradually transition from one to another category, e.g., the slats that gradually transition to slat track noise\deleted{, or the slat tones that vary from many dominant tones, see Figure~\ref{fig:Figure7} $c)$, to few weak tones}. Figure~\ref{fig:Figure8} also showed that a single source location can gradually or abruptly change \added{its mechanism} with increasing Reynolds number. \replaced[R1C01,R2C04h]{Some outliers, which are not close to their core group, are visible in Figure~\ref{fig:Figure11}. There are three possible explanations for this behavior. First, the manual labeling is wrong or ambiguous. Second, the automated feature calculation resulted in wrong values due to an insufficient definition or because the formula was not robust enough towards the degenerated spectra. Third, the spectra contained partially wrong information due to the limitations of the SIND process~\hbox{\citep{Goudarzi2020}}. The ability of UMAP to form separable source type groups highlights two results. First, the introduced feature space captures sufficient aeroacoustic information for the presented sources, under the condition that the labeling is correct. Thus, groups of multiple sources are formed in the UMAP based on the feature space. Second, the labeling is sufficiently good under the condition that the feature space correctly captures the sources' aeroacoustic behavior. Thus, the groups mainly contain unique source types. Since the source types were identified manually based on the source spectra and not on the introduced feature values, see Section~\ref{sec:classification}, this indicates that the label choices are reasonable.}{However, Figure~\ref{fig:Figure11} showed that the source groups are reasonable with respect to their members distribution in the introduced feature space.}\\

% The goal of the presented EDSS CRAFT was to help and guide experts to analyze new, complex data. Consequently, we used unsupervised learning techniques to identify source types with similar acoustical properties instead of supervised learning techniques. We chose HDBSCAN~\citep{Campello2013,McInnes2017} since it allowed us to cluster the data without any prior assumption of the expected number of source types or source distributions in the feature space. It also provided soft clustering, which estimated the probability for each source of belonging to each cluster. In combination with the mean feature values of each cluster which directly correspond to aeroacoustic properties and the cluster hierarchy, this should allow the expert to transparently analyze the clustering choices.\\

% clustering
We used HDBSCAN~\citep{Campello2013,McInnes2017} to cluster the sources without any prior assumption of the expected number of source types or source distributions in the feature space. For the Do728, \added[R1C36]{``clustering sources based on their aeroacoustic features''} (CRAFT) resulted in very intuitive clusters similar to our manual evaluation (see Figure~\ref{fig:Figure12}). As explained in the introduction this is a promising result since the feature space not only captures enough information to separate most of the source types, but the variance between the different source types is also greater than the variance of an unwanted phenomenon (e.g., a cluster for each angle of attack or each Reynolds number). \deleted[R1C01,R2C04h]{If the feature variance within a class became too large (e.g., the slat tracks or leading flap side edge), we often found multiple clusters that represent a single label class. This can be caused by differences in their spectra, which again result from different flow properties (e.g., the Reynolds number dependency in Figure~\ref{fig:Figure8}), geometric differences (e.g. two different types of flap tracks on the A320), source directivities (the slat tracks are observed from different radiation angles), or source identification artifacts (wrong labels). It can also be caused by the imbalance of label members within the dataset. }Since our source identification was often ambiguous and since not all clusters related to our labels the exact accuracy of CRAFT and the reason for the wrong cluster choices is up for debate. Our evaluation of the Do728 clustering was based on the confusion matrix in Figure~\ref{fig:Figure12} and resulted in an accuracy of $\SI{77.04}{\percent}$.\\

The usefulness of the chosen clusters can only be evaluated qualitatively, based on their consistency, their ability to separate source mechanisms from each other, and detecting sources for which the spatial location is misleading. For airframe noise, typical analyzed source regions are the whole slat, flap, flap side edge, and nacelle region, including the strakes~\citep{Ahlefeldt2013, Ahlefeldt2017}. Regarding the Do728 slat region, CRAFT showed that typical slat sources are distinctly different from slat track sources based on the cluster hierarchy. Occasionally, slat tones appeared with decreasing probability towards high Reynolds numbers. Often, they occurred at the slat positions, but at different angles of attack (slat noise appeared mainly at low angles of attack). CRAFT was able to separate these phenomena very well. However, multiple source types were clustered into the single cluster 14. Decreasing the sample size for \replaced{craft}{HDBSCAN} eventually results in a separation of these sources, but also numerous sub-clusters for the other source types.\\

For the A320, CRAFT's clusters did loosely correlate to our manually identified source types and thus, the accuracy of the result was much more difficult to evaluate. We encourage the readers to interpret the confusion matrix results based on their own experience. The evaluation based on our confusion matrix assessment in Figure~\ref{fig:Figure14} resulted in an accuracy of $\SI{61.52}{\percent}$. \replaced[R1C01,R2C04h]{Again, the usefulness of the chosen clusters is evaluated qualitatively. On the whole slat area, CRAFT correctly determined different clusters for the slat sources, the slat tracks, the inner slat, the slat tones, and the slat resonances. These were all sources with distinctly different spectra, similar to the Do728 spectra, see Figure~\ref{fig:Figure8}. When reviewing the source clusters 14, 15, and 16 we observed that all feature Strouhal number scaling tones. However, they can be separated in strong, high SPL tones in cluster 16, e.g. the high Reynolds number spectrum in Figure~\ref{fig:Figure11} $b)$, and weaker tones in cluster 15, e.g. the low Reynolds number spectra in Figure~\ref{fig:Figure8} $b)$, and weak tones in cluster 14. When reviewing the soft cluster confidences for sources in cluster number 14, they all showed high confidence for cluster number fifteen and vice versa. Thus, CRAFT predicted reasonable sub-categories for Strouhal number scaling tones. However, based on this observation the clustering of the nacelle track, flap gap, and wing tip is not meaningful, since these sources feature Helmholtz number scaling tones.}{The lower clustering accuracy can be accounted to the larger feature variance compared to the Do728 sources, as depicted in Figure~\ref{fig:Figure9}. This is mainly caused by the worse quality of the spectra, due to the smaller microphone array employed for the measurements.}\\

% Most of the clusters proved to be meaningful representations of the source type categories identified by the authors. While the underlying physical mechanisms, the correct labels and thus, CRAFT's accuracy is up for debate, the clustering was mostly consistent between the Do728 and A320 dataset, which featured similar source types. We showed that CRAFT worked on the very large Do728 dataset with 928 individual sources, consisting of six spectra each, and the relatively small A320 dataset with 408 sources, consisting of three spectra each. 

\section{Conclusion}
The goal of this paper was to use clustering to group multiple sources based on their aeroacoustic properties to reveal underlying physical mechanisms and guide the acoustic expert in identifying and analyzing the sources correctly. The general assumption of this Expert Decision Support System ``CRAFT'' was that the physical mechanism of a source can be determined by the change of its properties over the Mach number. Thus, source measurements at multiple flow speeds are required. Example data of Dornier 728 and Airbus A320 CLEAN-SC beamforming maps were used at different flow speeds, angles of attack, and Reynolds numbers to derive the aeroacoustic properties and employ the presented method.\\

To reduce the complex acoustic properties to a data space that can also be understood by the machine, we introduced a feature-set that expresses these properties as a combination of single, numerical values. These features are independent or averaged over the spectra at different Mach numbers, which enables CRAFT to compare and cluster sources from different experiments. To further evaluate the data, we presented exemplary source spectra and corresponding manual labeling of the sources. We showed that the manual labeling of the sources is often ambiguous due to degenerated spectra, multiple source mechanisms, or Reynolds number dependent trends which resulted in additional uncertainty. Despite the ambiguous manual source type identification, many source types formed distinguishable distributions in the introduced feature space, which was visualized with a Uniform Manifold Approximation and Projection for Dimension Reduction.\\

We used Hierarchical Density Based Clustering for Applications with Noise to group the sources in the introduced feature space which did not include meta-information such as the source position, the angle of attack, or the Reynolds number. The clusters corresponded mostly to the manually identified source types were consistent between the datasets and provided the necessary information to identify sources that behaved atypically for their spatial locations. For example, it allowed the correct identification of multiple source types that were all located on a single slat position. The result also provided a confidence estimation for the clustering results. We showed that sources are mostly clustered wrongly by CRAFT at low confidence, while the clustering with high confidence is mostly correct. Experts can discard predictions below a confidence threshold which further increases the prediction accuracy.\\

% Being able to quickly analyze sources and build labeled datasets this way has many advantages. It enables the reliable analysis of source type-dependent features such as the power scaling with a mean and a standard deviation. Since the presented features are independent of the absolute Mach number, CRAFT also enables the comparison of different models measured in different wind tunnels at different flow configurations. Since the analysis is run within minutes instead of typically weeks or months for datasets of the presented sizes, CRAFT can also be used to detect spurious sources in-situ during experiments.\\

For future work, we plan to analyze more data with the introduced method to further evaluate the quality and reliability of the proposed features. We also hope to start a discussion in the aeroacoustic community on the selected properties and their corresponding features, possible shortcomings, alternative definitions or calculation methods, and the manual identification of aeroacoustic source phenomena.

\section*{Acknowledgments}
We want to thank the experts of the aeroacoustic group G\"ottingen, especially Dr. Thomas Ahlefeldt, for the helpful discussions on the analyzed beamforming results. The authors also acknowledge the DLR, Institute of Aerodynamics and Flow Technology, Department of Experimental Methods (contact: C. Spehr) for providing the SAGAS software which generated the beamforming and CLEAN-SC results for this paper. We also thank the reviewers for their helpful comments and insights.

\bibliography{main}{}

\end{document}